\documentclass[runningheads]{llncs}

\usepackage[T1]{fontenc}
\usepackage{graphicx}
\usepackage{booktabs}
\usepackage{tabularx}
\usepackage{array}
\usepackage{multirow}
\usepackage{longtable}
\usepackage{placeins}
\usepackage{amsmath,amssymb}
\usepackage{pifont}
\usepackage[table]{xcolor}
\usepackage{tikz}
\usetikzlibrary{positioning,shapes.geometric,arrows.meta,fit,backgrounds,calc}
\usepackage{url}
\usepackage[hidelinks]{hyperref}
\usepackage{orcidlink}

\newcommand{\cmark}{\ding{51}}
\newcommand{\xmark}{\ding{55}}
\newcolumntype{Y}{>{\raggedright\arraybackslash}X}

\definecolor{cellStrong}{RGB}{215,238,220}
\definecolor{cellPartial}{RGB}{253,226,180}
\definecolor{cellWeak}{RGB}{250,219,216}
\newcommand{\gr}{\cellcolor{cellStrong}\cmark}
\newcommand{\pa}{\cellcolor{cellPartial}$\circ$}
\newcommand{\ba}{\cellcolor{cellWeak}\xmark}
\newcommand{\na}{\textsc{n/a}}                    % not applicable

\renewcommand{\tablename}{Table}
\renewcommand{\figurename}{Figure}

\title{SoK: Cryptocurrency Mixing and Anonymity -- Architectures, Threat Models, Operational Aspects and Security}

\titlerunning{SoK: Cryptocurrency Mixing and Anonymity}

\author{Juraj Mariani\inst{1}\texorpdfstring{\,\orcidlink{0009-0003-0942-0634}}{} \and
	Martin Perešíni\inst{1,3}\texorpdfstring{\,\orcidlink{0000-0002-2875-9567}}{} \and
	Ivan Homoliak\inst{1,2}\texorpdfstring{\,\orcidlink{0000-0002-0790-0875}}{}}
\authorrunning{J. Mariani et al.}
\institute{Brno University of Technology, Faculty of Information Technology, Božetěchova 1/2, 612 00 Brno, Czechia\\	
	\and
	Slovak University of Technology in Bratislava, Faculty of Informatics and Information Technologies, Ilkovičova 2, 842 16 Bratislava, Slovakia\\
    \and
	Communication Systems Group CSG, Department of Informatics IfI, University of Zürich UZH, Zürich, Switzerland\\
    \email{\{imariani,iperesini,homoliak\}@fit.vut.cz}}
\begin{document}
\maketitle

\begin{abstract}
Public blockchains record transaction histories that enable address clustering, taint analysis, and cross-service attribution, thereby motivating the development of mixers and privacy layers.
Our work presents a structured scoping review of 22 representative systems, defining a common unlinkability objective and five adversary arche\-types.
We evaluate these systems against a taxonomy of attack surfaces, including chain analysis, timing inference, custodial compromise, coordination abuse, network metadata, and trusted execution compromise.
%In theory, nominal anonymity set size and cryptographic strength specify a system's privacy.
%In practice, effective anony\-mity is governed by operational factors, including transaction denominations, the presence of cover traffic, relayer network behavior, and the design of compliance interfaces.
While nominal anonymity-set size and cryptographic strength characterize privacy in theory, effective anonymity in practice depends on transaction denominations, cover traffic, relayer behavior, and compliance-interface design.
%Our analysis distinguishes nominal from effective anonymity and extracts four core lessons: 
Distinguishing nominal from effective anonymity, we derive four core lessons:
(1) Privacy is strongest when integrated into everyday transactions, since standalone mixing creates an easily profiled user subset; 
(2) Trust points, including operators, peer quorums, and hardware enclaves, must be explicit so users know who can break privacy;
(3) Network metadata, including gas funding and timing, must be treated formally as protocol data in privacy evaluations; and
(4) Compliance should use auditable cryptographic predicates for selective disclosure rather than broad operator discretion.
Ultimately, our systematization clarifies the strengths, failures, and future requirements of blockchain privacy architectures.

\end{abstract}

\section{Introduction}
\label{sec:intro}

Blockchain systems provide integrity and auditability by making transaction histories globally visible.
This transparency supports consensus and verification, but it also enables address clustering, taint analysis, behavioral profiling, and attribution across services~\cite{moserTaint}.
Users of public chains therefore receive pseudonymity rather than anonymity by default.
Its strength depends on the difficulty of linking ledger events to each other and to real identities.
Transaction privacy is consequently important for applications that must protect balances, counterparties, or spending patterns while retaining public verifiability.

Cryptocurrency mixing emerged as the first response to that gap.
In its narrowest sense, a mixer is an intermediary or protocol that obscures the relationship between a user's input and output transactions.
The design space is broader in practice.
Some systems rely on custodial pooling and redistribution.
Others create ambiguity through joint transactions, encode withdrawal rights in private notes, or internalize privacy in the base protocol through ring signatures, shielded transactions, or confidential execution.
The literature is therefore fragmented.
Mixing can denote a service category or a more general unlinkability mechanism.

That fragmentation matters because privacy claims are often made for a cryptographic mechanism rather than for the complete deployed system.
The central premise of this work is that the gap between cryptographic and operational privacy drives many mixer failures.
A large advertised anonymity set need not yield a large effective set once timing, amount patterns, relayer behavior, network metadata, operator records, or hardware assumptions are considered.
Mixer forensics and smart contract analysis show that practical deanonymization often exploits this operational context rather than the core primitive~\cite{howToSpotAMixer,TCAnalysis,Wang2023ZKPmixers,TIPPE2025301876}.
These observations motivate three research questions that structure the review.

\paragraph{\textbf{Research Questions.}}
\label{sec:rqs}
We specify the following research questions:
% \begin{itemize}
%     \item[\textbf{RQ1.}] How can heterogeneous cryptocurrency mixing systems be organized and compared under a common transaction unlinkability objective and adversary model?
%     \item[\textbf{RQ2.}] Which adversary archetypes remain effective under realistic conditions, and which operational factors most strongly determine the gap between nominal anonymity (the theoretical size of the user pool) and effective anonymity (the actual indistinguishability achieved in practice)?
%     \item[\textbf{RQ3.}] Which design principles, evaluation practices, and open problems follow from the taxonomy, the adversary-driven comparison, and the matrix of attack surfaces?
% \end{itemize}

\noindent\textbf{RQ1.} How can heterogeneous cryptocurrency mixing systems be organized and compared under a common transaction unlinkability objective and adversary model?

\noindent\textbf{RQ2.} Which adversary archetypes remain effective under realistic conditions, and which operational factors most strongly determine the gap between nominal anonymity (the theoretical size of the user pool) and effective anonymity (the actual indistinguishability achieved in practice)?

\noindent\textbf{RQ3.} Which design principles, evaluation practices, and open problems follow from the taxonomy, the adversary-driven comparison, and the matrix of attack surfaces?

\paragraph{\textbf{Design Space.}}
We organize the design space around one guiding question: which actor can still reconstruct the transaction graph once the surrounding system is accounted for.
The taxonomy addresses RQ1.
The comparative review and the matrix of attack surfaces address RQ2.
The synthesis addresses RQ3.

This work studies cryptocurrency mixing as a systems problem rather than only as a cryptographic feature.
It explains why mixer families fail in practice, where their trust assumptions reside, and how implementation details reshape the privacy promised by a protocol.
Existing surveys focus on Bitcoin and Ethereum mixer catalogues~\cite{Arbabi2025bitcoinmixing} or address privacy-preserving payments more broadly~\cite{Nardelli2025paymentsurvey}.
We instead treat mixing as a blockchain privacy problem that spans architectural families.

\paragraph{\textbf{Contributions.}}
The main contributions are as follows.
\begin{itemize}
    \item We define the scope across seven architectural families and position it against recent mixer and payment surveys.
    We formalize transaction unlinkability from the adversary's perspective and introduce five adversary archetypes, covering chain analysis, custodial compromise, active coordination, network observation, and hardware compromise.
    \item We review 22 representative systems and code each by trust anchor, source of cover traffic, observable metadata, and dominant failure mode.
    Then we compared attack surfaces under an explicit rubric (\autoref{tab:attacks}) and use Tornado Cash, Monero, and RAILGUN to distinguish heuristic candidate counts from effective anonymity.
    \item We derive design lessons for blockchain privacy systems, including metrics that reflect known attacks, accountable coordination, reduced metadata leakage, narrowly defined compliance mechanisms, and programmable private execution.
\end{itemize}

The remainder of this work answers the research questions in turn.
\autoref{sec:method} defines the scope and methodology.
\autoref{sec:threat} presents the unlinkability objective and five adversary archetypes.
\autoref{sec:taxonomy} addresses RQ1.
\autoref{sec:systems}, the comparison tables, and the matrix of attack surfaces address RQ2.
\autoref{sec:synthesis} and the conclusion address RQ3.

\section{Scope, Methodology, and Positioning}
\label{sec:method}

This work is a structured scoping review rather than a narrow protocol survey.
Its scope includes systems designed to reduce the linkability between asset ingress and egress events on a public ledger.
It covers custodial services, coordinated Bitcoin protocols, account based contract mixers, privacy layers, and privacy native assets.
It also considers exchange mediated and cross chain obfuscation when a mechanism is used to interrupt tracing across ledgers, although this class does not resist the exchange operator.
Mix networks in the tradition of Chaum~\cite{mixerHistory} are outside the scope because they protect messages rather than ledger asset events.
Network obfuscation, such as Dandelion++~\cite{dandelionpp}, is considered only when combined with an on-chain mixing architecture.

\subsection{Search Strategy and Corpus Construction}

We constructed the corpus iteratively up to 10 July 2026.
Searches covered IEEE Xplore, the ACM Digital Library, SpringerLink, USENIX, FC, and WPES.
For deployed systems without an archival publication, we consulted authoritative technical artifacts, such as protocol specifications, technical reports, source repositories, operator reports, and forensic analyses.
We complemented database searches with backward and forward citation tracing from relevant surveys and empirical studies.

Searches combined terms for mixing and privacy mechanisms with concrete system names and forensic terms.
Representative terms included ``cryptocurrency mixer,'' ``CoinJoin,'' ``Tornado Cash,'' ``shielded transactions,'' ``privacy layer,'' ``zkSNARK,'' ``unlinkability,'' ``clustering,'' ``deanonymization,'' and ``tracing,'' with additional queries for Wasabi, RAILGUN, Obscuro, and Monero.
A source was included when it specified a mixing or unlinkability mechanism, described an implemented privacy system, analyzed the traceability of a deployed mixer, or reported reproducible evidence on a recurring operational or implementation failure.
We excluded sources that address regulatory policy without technical mechanism analysis, duplicate descriptions when a more authoritative source was available, generic graph analysis unrelated to a mixer architecture, and purely theoretical proposals without implementation, deployment, or empirical data.
The unit of analysis is a deployed system or protocol family rather than an individual publication.

The selection process is summarized in \autoref{fig:method}.
The final corpus comprises 22 representative systems in seven families: custodial mixers, coordinated protocols, contract mixers, privacy layers, privacy native currencies, confidential execution platforms, and exchange mediated obfuscation.
The selection prioritizes technical depth, evidence of implementation or deployment, and coverage of distinct design families over exhaustive enumeration.
TumbleBit~\cite{tumblebit}, Penumbra~\cite{penumbra}, and Namada~\cite{namada} appear only as contextual comparators and are not coded as part of the 22-system corpus (they are mentioned to contrast specific architectural features like payment hubs and shielded pools).

\begin{figure}[t]
\centering
\resizebox{0.78\textwidth}{!}{%
\begin{tikzpicture}[
  node distance=4mm and 7mm,
  stage/.style={draw, rounded corners, align=left, inner sep=3pt, font=\scriptsize, text width=5.8cm, fill=blue!5, minimum height=0.9cm},
  excl/.style={draw, dashed, rounded corners, align=left, inner sep=3pt, font=\scriptsize, text width=4.2cm, fill=black!2},
  arr/.style={-{Latex[length=1.5mm]}, semithick},
]
\node[stage] (id) {\textbf{Search and retrieval}\\
Search indexed databases, security venues, system names, and reference lists.
Collect primary technical artifacts.};
\node[stage, below=of id] (sc) {\textbf{Screening}\\
Remove duplicate or irrelevant material.
Screen titles, abstracts, keywords, and source authority.};
\node[excl, right=of sc, yshift=0mm] (ex1) {\textbf{Excluded at screening:} regulatory policy without mechanism analysis, generic graph analysis without a mixer architecture, and duplicate coverage.};
\node[stage, below=of sc] (el) {\textbf{Eligibility assessment}\\
Assess mechanism detail, implementation or deployment evidence, and relevance to practical traceability.};
\node[excl, right=of el, yshift=0mm] (ex2) {\textbf{Excluded at eligibility:} theoretical proposals without deployment evidence and sources with insufficient mechanism detail.};
\node[stage, below=of el, fill=green!6] (in) {\textbf{Included corpus}\\
22 representative systems in seven architectural families.
Served as an unit of analysis of deployed system or protocol family.};
\draw[arr] (id) -- (sc);
\draw[arr] (sc) -- (el);
\draw[arr] (el) -- (in);
\draw[arr, dashed] (sc.east) -- (ex1.west);
\draw[arr, dashed] (el.east) -- (ex2.west);
\end{tikzpicture}}
\caption{Review protocol for corpus construction.
The process adapts scoping review stages to deployed systems rather than primary studies.
It records the search, screening, eligibility, and inclusion decisions that yielded the corpus of 22 systems. }
%\ih{It would be good to mention number of papers after each filtering stage.}
\label{fig:method}
\end{figure}

\subsection{Meaning of Mixing}

We use mixing in a functional sense: a system qualifies when it intentionally increases uncertainty about the correspondence between a source and destination asset event.
Privacy-preserving cryptocurrencies and confidential execution platforms are legitimate comparison targets even when they are not mixers in the strict service sense, because they solve the same unlinkability problem at a different layer.
Centralized exchanges create ambiguity for external observers without offering privacy against the exchange itself.
They differ both in mechanism and in the entity that can collapse uncertainty.

\subsection{Coding Protocol and Validity Threats}

For each retained system we recorded the ledger and deployment model, privacy primitive, trust anchor, source of cover traffic, observable metadata, and dominant failure modes.
We used peer-reviewed empirical studies where available, followed by primary specifications and official documentation.
Each cell in the matrix of attack surfaces (\autoref{tab:attacks}) applies the column-specific criteria in \autoref{sec:rubric}.
The appendix records the evidence for non-trivial assignments.

Three validity threats apply.
First, documentation quality is uneven, especially for defunct or seized custodial services known mainly through forensic reconstruction.
Second, several recent systems have limited longitudinal evidence, so claims about them are primarily architectural.
Third, legal and operational conditions can change faster than protocol designs.
% Policy and enforcement statements were checked against primary sources and are current to 10 July 2026.
The technical comparisons remain meaningful even if particular sanctions or obligations evolve.

\subsection{Evaluation Rubric}
\label{sec:rubric}

The decision criteria for the attack surfaces are listed in \autoref{tab:rubric}. These make the assignments in \autoref{tab:attacks} explicit and reproducible.
The matrix records six technical surfaces.
They map to the archetypes in \autoref{sec:archetypes}: operator compromise to $\mathcal{A}_{\mathrm{custodial}}$; timing and amount linkage to $\mathcal{A}_{\mathrm{chain}}$; Sybil disruption to $\mathcal{A}_{\mathrm{coord}}$; network metadata to $\mathcal{A}_{\mathrm{net}}$; and TEE compromise to $\mathcal{A}_{\mathrm{hw}}$.
Legal compulsion that discloses operator records is included under operator compromise.
Sanctions, delisting, and other availability effects are analyzed separately because they do not by themselves reveal a transaction mapping.

An assignment of \cmark{} requires a structural defense under the model in \autoref{sec:threat}, with no documented counterexample in a deployed instance.
$\circ$ denotes conditional resistance, for example when the outcome depends on behavior, participation volume, optional protections, or a contested assumption.
\xmark{} denotes no structural defense or a documented successful attack.
\na{} denotes an inapplicable surface, such as TEE attacks against a system without an enclave.
The marks are ordinal assessments supported by the cited evidence, not quantitative measurements.
The appendix gives the expanded criteria and evidence notes.

\begin{table}[t]
\centering
\footnotesize
\setlength{\tabcolsep}{3pt}
\renewcommand{\arraystretch}{1.08}
\caption{Decision criteria for the matrix of attack surfaces.}
\label{tab:rubric}
\begin{tabular}{@{}>{\raggedright\arraybackslash}m{1.85cm}>{\raggedright\arraybackslash}m{3.22cm}>{\raggedright\arraybackslash}m{3.22cm}>{\raggedright\arraybackslash}m{3.22cm}@{}}
\toprule
\textbf{Surface} & \textbf{\cmark{} Resistant} & \textbf{$\circ$ Partial} & \textbf{\xmark{} Exposed} \\
\midrule
\textbf{Operator compromise} & No actor holds the mapping & Blinded view or logging risk & Custody or records reveal the mapping \\
\arrayrulecolor{black!20}\specialrule{0.25pt}{0.8pt}{0.8pt}\arrayrulecolor{black}
\textbf{Timing analysis} & Mandatory protection defeats correlation & Depends on volume or user behavior & Empirical linkage succeeds \\
\arrayrulecolor{black!20}\specialrule{0.25pt}{0.8pt}{0.8pt}\arrayrulecolor{black}
\textbf{Sybil disruption} & Noninteractive or robust admission & Depends on peers or incentives & Sybil aborts can collapse rounds \\
\arrayrulecolor{black!20}\specialrule{0.25pt}{0.8pt}{0.8pt}\arrayrulecolor{black}
\textbf{Amount linkage} & Amounts are hidden or normalized & Partial matching partitions candidates & Visible amounts enable linkage \\
\arrayrulecolor{black!20}\specialrule{0.25pt}{0.8pt}{0.8pt}\arrayrulecolor{black}
\textbf{TEE compromise} & Privacy survives enclave compromise & Compromise reveals only part of the mapping & Compromise reveals protected state \\
\arrayrulecolor{black!20}\specialrule{0.25pt}{0.8pt}{0.8pt}\arrayrulecolor{black}
\textbf{Network metadata} & Mandatory decoupling of network identity & Optional or external tools reduce exposure & No specified network decoupling \\
\bottomrule
\end{tabular}
\end{table}

The rubric remains qualitative.
Readers may reasonably place a borderline cell at an adjacent level when only architectural arguments exist, and new attacks or measurements may justify reassessment.
Future work should supplement such ordinal comparisons with bounds on effective anonymity under a fixed suite of heuristics.

\subsection{Positioning Against Related Work}

Recent surveys document the mixing ecosystem at different levels of granularity.
Arbabi et al.~\cite{Arbabi2025bitcoinmixing} review Bitcoin and Ethereum mixing services, while Nardelli et al.~\cite{Nardelli2025paymentsurvey} survey privacy-preserving payments across crypto assets and central bank digital currencies.
Forensic studies show that deployed services can still be detected or partially traced~\cite{howToSpotAMixer,TIPPE2025301876,KumaresanTracingSurvey2025}.
Other work studies zero-knowledge mixer claims~\cite{Wang2023ZKPmixers}, Tornado Cash clustering~\cite{TCAnalysis}, custodial economics~\cite{MixerServicesSecurityAnalysis}, and cross-ledger tracing~\cite{crossChainSecurityAnalysis}.
At the broader ecosystem level, the Security Reference Architecture (SRA)~\cite{sra} established a standardized, layered model for evaluating general blockchain security threats.
Our survey extends this systematization approach by contributing an evaluation framework specialized for anonymity layers. 
It identifies where privacy is generated, which assumptions support it, which adversaries remain effective, and how operational details reduce the privacy suggested by protocol primitives (\autoref{tab:related-surveys}, appendix).

\section{System Model and Adversary Model}
\label{sec:threat}

To compare heterogeneous systems meaningfully, we treat mixing as a privacy objective over observable ledger events rather than as a specific protocol family.

\subsection{Terminology}
\label{sec:terminology}
We adopt the terminology of Pfitzmann and Hansen~\cite{pfitzmann2010terminology}.
Two items of interest (IOIs), such as a deposit and a withdrawal, are unlinkable from an adversary's perspective if the adversary cannot sufficiently distinguish whether they are related.
The anonymity set of an observed IOI is the set of alternatives consistent with that adversary's observations and knowledge.
Transaction unlinkability is the special case in which the two IOIs are an ingress event and its true egress event.
Its candidate set is defined below.
We summarize uncertainty with a probability distribution over candidates rather than a raw count~\cite{serjantov2003infotheoretic,diaz2003measuring}.

\subsection{Unlinkability Objective and Effective Anonymity}
Let $d_i$ denote a deposit, spend, or ingress event.
For an instantiated adversary $\mathcal{A}$ and a fixed realized view $v$, let $W_i(v)$ be the effective anonymity set.
It contains the candidate egress events consistent with the protocol rules and all evidence in $v$.
The view can include ledger data, timing, amounts, protocol messages, relayer interactions, operator records, or other side information allowed by the model.
A mixing system seeks to limit the adversary's ability to link $d_i$ to the correct egress event $w_i \in W_i(v)$.
The advantage is defined by
\begin{equation}
\mathrm{Adv}_{\mathrm{link}}(\mathcal{A};v) =
\Pr[\mathcal{A}(d_i,v) = w_i \mid V_{\mathcal{A}}=v]
- \frac{1}{|W_i(v)|}.
\label{eq:adv}
\end{equation}
The conditional probability is taken over protocol randomness, the hidden input and output relation consistent with $v$, and the adversary's internal randomness.
The second term is a uniform cardinality baseline for the realized candidate set.
It is a reference point rather than an assertion that the posterior is uniform.

Equation~\eqref{eq:adv} adapts the cryptographic convention of measuring success above a stated baseline~\cite{katz2020introduction}.
Equation~\eqref{eq:keff} applies the established anonymity measurement principle that the posterior distribution, not only its support, determines the remaining uncertainty~\cite{serjantov2003infotheoretic,diaz2003measuring,clauss2006structuring}.
To make the probability distribution explicit, let $p_w=\Pr[w=w_i\mid V_{\mathcal{A}}=v]$.
It is defined over $W_i(v)$, with $\sum_{w\in W_i(v)}p_w=1$ and $p_w\geq 0$.
Anonymity research commonly evaluates the posterior rather than only its support size~\cite{serjantov2003infotheoretic,diaz2003measuring,clauss2006structuring}.
We use the reciprocal of collision probability, equivalently the exponential of R\'enyi entropy of order two, as an effective candidate count:
\begin{equation}
k_{\mathrm{eff}}(v) = \frac{1}{\sum_{w \in W_i(v)} p_w^2}.
\label{eq:keff}
\end{equation}
The value $k_{\mathrm{eff}}(v)$ equals $|W_i(v)|$ only when the posterior is uniform.
It is smaller when probability mass concentrates on fewer candidates.

The expressions adapt established tools and are not claimed as novel.
The contribution is their consistent application across mixer families.
Evaluating $\mathrm{Adv}_{\mathrm{link}}(\mathcal{A};v)$ and $k_{\mathrm{eff}}(v)$ under the same five archetypes and evidence classes yields comparable statements for custodial services, coordinated protocols, contract mixers, privacy layers, and privacy native ledgers.
This distinction matters when public interfaces advertise large pools but practical heuristics narrow the plausible candidates~\cite{TCAnalysis,Wang2023ZKPmixers}.
The advertised pool size is not itself $W_i(v)$.
The candidate set remains after filtering by amount structure, timing rules, coordination metadata, cross chain conversion rates, visible operator records, and other deployment side channels.

\subsection{Evaluation Criteria}

A mixing design is robust only when it addresses more than unlinkability in isolation.
We evaluate five dimensions.
\textbf{(1) Transaction unlinkability} concerns the input and output relation under the adversary model.
\textbf{(2) Amount and destination privacy} concerns the values and recipients that remain visible.
\textbf{(3) Liveness and disruption tolerance} concerns completion under churn, aborts, and low participation.
\textbf{(4) Trust minimization} identifies which operator, coordinator, relayer, validator, exchange, or hardware vendor can reconstruct the linkage.
\textbf{(5) Deployability} covers realistic fees, latency, liquidity, and regulatory constraints.

\subsection{Adversarial Goals}

Across mixer families, the adversary pursues one or more related goals.
Input and output linkage identifies which output belongs to which input.
Entity linkage clusters addresses, notes, or accounts into the same user.
Amount linkage narrows candidates through denominations, change, or conversions.
Timing and network inference exploits intervals between deposit and withdrawal, message timing, and IP metadata.
Anonymity set degradation and service denial reduce the cover set through Sybil participation~\cite{sybilAttack}, targeted DoS attacks~\cite{poscopor}, censorship, or aborts.
These goals bear on architectures differently, a mapping made precise by the matrix of attack surfaces (\autoref{tab:attacks}).

\subsection{Adversarial Knowledge and Capabilities}
\label{sec:capabilities}

We adopt a Kerckhoffs baseline: the adversary knows the protocol, primitives, contract code, and ledger history.
Depending on the scenario, it can also observe or control operator records, protocol participation, network metadata, or trusted hardware.
Known TEE attacks include side channels, rollback, and memory attacks~\cite{IntelSGX,SGXattack,brasser2017dr,biondo2018guard,matetic2017rote,cloosters2020teerex,Borrello2022AEPIC}.

\subsection{Adversary Archetypes}
\label{sec:archetypes}

\autoref{tab:archetypes} defines five archetypes used in the taxonomy, the system review, and the attack matrix.
They are neither exhaustive nor mutually exclusive.
A realistic adversary can compose them, for example $\mathcal{A}_{\mathrm{chain}} + \mathcal{A}_{\mathrm{net}}$ against a contract mixer.
We flag such compositions when they dominate a system's effective privacy.

\begin{table}[t]
\centering
\footnotesize
\setlength{\tabcolsep}{3pt}
\renewcommand{\arraystretch}{1.08}
\caption{Adversary archetypes and their relation to the attack matrix.}
\label{tab:archetypes}
\begin{tabular}{@{}>{\raggedright\arraybackslash}m{1.7cm}>{\raggedright\arraybackslash}m{5.0cm}>{\raggedright\arraybackslash}m{5.0cm}@{}}
\toprule
\textbf{Adversary} & \textbf{View and actions} & \textbf{Goal and attack surface} \\
\midrule
$\mathcal{A}_{\mathrm{chain}}$ & Observes the public ledger, contract interactions, and relayer transactions~\cite{moserTaint,TCAnalysis,beres2021}. & Infers input, output, and amount links; timing and amount surfaces. \\
$\mathcal{A}_{\mathrm{custodial}}$ & Obtains operator or exchange records through compromise, seizure, insider access, or legal compulsion. & Reconstructs the mapping; operator compromise. \\
$\mathcal{A}_{\mathrm{coord}}$ & Joins rounds, creates Sybil identities~\cite{sybilAttack}, aborts, or manipulates admission. & Reduces cover traffic or prevents completion; Sybil disruption. \\
$\mathcal{A}_{\mathrm{net}}$ & Observes IP metadata, propagation timing, and relayer fingerprints~\cite{dandelionpp,Tramer2020RemoteSideChannel}. & Links users to transactions; network metadata. \\
$\mathcal{A}_{\mathrm{hw}}$ & Exploits TEE side channels, rollback, or microarchitectural leakage. & Recovers protected execution state; TEE attacks. \\
\bottomrule
\end{tabular}
\end{table}

\subsection{Assumptions and Limitations}

Unless stated otherwise, we assume standard cryptographic primitives are not broken.
The adversary does not invert hashes, forge zero-knowledge proofs, or break signature schemes directly.
We do not, however, assume that surrounding systems are ideal.
Users may withdraw too quickly, reuse addresses, choose rare denominations, leak metadata to relayers, or rely on centralized operators.
Private execution platforms may inherit weaknesses from the hardware they trust.
These assumptions outside the cryptographic core are often decisive.

\section{Taxonomy of Mixing Architectures and Privacy Primitives}
\label{sec:taxonomy}

We code each system as a tuple $(L,P,R)$.
The deployment locus $L$ is a custodial service, coordinated protocol, contract, privacy layer, base ledger, confidential execution platform, or exchange flow.
The primary privacy primitive $P$ is pooling, joint construction, a zero knowledge proof, a ring signature, a confidential transaction, or trusted execution.
The reconstruction authority $R$ identifies the actor or compromised component that can recover the mapping.
\autoref{sec:appendix-coding} reports the tuple for every system.

The family is assigned by $L$.
The fields $P$ and $R$ resolve hybrids and state their residual trust explicitly.
For example, Obscuro~\cite{obscuroPaper} remains a custodial service by locus, but its reconstruction authority is enclave compromise rather than the service operator under the stated hardware assumptions.
Coordinated protocols require sufficiently honest and available participants.
Contract mixers and privacy layers expose host ledger metadata.
Privacy native currencies place unlinkability in the base protocol.
Confidential execution platforms rely on trusted hardware.
\autoref{fig:taxonomy} depicts this rule.
\autoref{sec:primitives} discusses the primitives.

\begin{figure}[t]
\centering
\begin{tikzpicture}[
  root/.style={draw, rounded corners, fill=black!8, align=center, inner sep=3pt, text width=2.4cm, font=\footnotesize},
  branch/.style={draw, rounded corners, align=center, inner sep=2pt, text width=2.2cm, font=\scriptsize\bfseries},
  fam/.style={draw, rounded corners, align=left, inner sep=2pt, text width=4.6cm, font=\scriptsize},
  arr/.style={-{Latex[length=1.4mm]}, semithick},
]
\node[root] (r) {Cryptocurrency mixing};
\node[branch, above right=5mm and 7mm of r] (nr) {Protocol or ledger locus};
\node[branch, below right=5mm and 7mm of r] (cust) {Service or exchange locus};

\node[fam, right=5mm of nr, yshift=15mm] (f1) {\textbf{Coordinated:} users jointly construct a transaction};
\node[fam, below=0.8mm of f1] (f2) {\textbf{Contract:} withdrawal authority is cryptographically hidden};
\node[fam, below=0.8mm of f2] (f3) {\textbf{Privacy layer:} a shielded pool or private execution settles to a host chain};
\node[fam, below=0.8mm of f3] (f4) {\textbf{Privacy native:} unlinkability is part of the base protocol};
\node[fam, below=0.8mm of f4] (f5) {\textbf{Confidential execution:} protected state resides in trusted hardware};

\node[fam, right=5mm of cust, yshift=-1mm] (g1) {\textbf{Custodial:} a service receives and redistributes funds};
\node[fam, below=0.8mm of g1] (g2) {\textbf{Exchange mediated:} custody and conversion span asset ledgers};

\draw[arr] (r) -- (nr);
\draw[arr] (r) -- (cust);
\draw[arr] (nr.east) -- (f1.west);
\draw[arr] (nr.east) -- (f2.west);
\draw[arr] (nr.east) -- (f3.west);
\draw[arr] (nr.east) -- (f4.west);
\draw[arr] (nr.east) -- (f5.west);
\draw[arr] (cust.east) -- (g1.west);
\draw[arr] (cust.east) -- (g2.west);
\end{tikzpicture}
\caption{Taxonomy by deployment locus.
The tuple $(L,P,R)$ supplements each family label with its privacy primitive and reconstruction authority.
Representative systems appear in \autoref{tab:comparison}.}
\label{fig:taxonomy}
\end{figure}

\subsection{Architectural classes}

We summarize each class here and defer the detailed system analysis to \autoref{sec:systems}.

\subsubsection{Custodial Mixers.}
Custodial mixers accept user funds into operator-controlled addresses and later redistribute them from pooled reserves.
They need little coordination, but the operator, or anyone who later compromises it, can reconstruct the true mapping.
The pooling and redistribution also leave recognizable graph motifs such as batching and peeling chains~\cite{moneyLaunderingInquiry,SecureSystems,TIPPE2025301876}.

\subsubsection{Coordinated Bitcoin Mixers.}
CoinJoin, CoinShuffle, CoinParty, and Wasabi Wallet keep funds under user control until participants construct a joint transaction~\cite{CJoin,coinShuffle,cParty,wasabiwal}.
Equal outputs create ambiguity for a passive observer.
The class removes custody but requires interactive liveness.
Its privacy therefore depends on Sybil resistance~\cite{sybilAttack}, accountability, and sufficient compatible participants.

\subsubsection{Contract Mixers.}
Account-based platforms can hold privacy state in a contract.
Users deposit a fixed denomination, receive a note, and later prove a right to withdraw without revealing the deposit they own~\cite{AMR,mixeth,TornadoCash}.
Succinct arguments originate with Bitansky et al.~\cite{zk-SNARKs} and were applied to anonymous currency in Zerocoin~\cite{Zerocoin}.
The design removes custody and is auditable, but deposits, gas funding, contract interactions, and relayer patterns remain visible.
Effective anonymity thus depends on behavior as well as the cryptographic primitive~\cite{TCAnalysis,Wang2023ZKPmixers}.

\subsubsection{Privacy Layers.}
Privacy layers embed private state in an application protocol above a contract platform.
RAILGUN maintains a shielded UTXO pool for arbitrary tokens and variable amounts~\cite{railgun}.
Aztec is a Layer 2 rollup in which private state transitions execute on the client and settle on chain with PLONK proofs~\cite{aztec,plonk}.
These designs support programmable private computation but inherit host chain constraints and require sufficient cover traffic. Additionally, zkSNARKs can protect complex DeFi instruments, such as Hopinzero's on-chain options, by hiding commercially sensitive terms~\cite{kostal2026hopinzero}.

\subsubsection{Privacy Native Chains and Confidential Execution.}
Some systems move unlinkability into the base protocol.
Monero uses ring signatures, stealth addresses, and confidential transactions.
Zcash uses shielded transactions backed by zkSNARKs.
Grin and Beam implement Mimblewimble variants~\cite{Monero,RingCT,zcash,mimblewimble}.
Oasis, Secret, Integritee, and Phala use confidential computation in trusted hardware~\cite{oasis,secret,integritee,phala}.
These systems show that privacy can be a base protocol property rather than a service overlay, a principle also applied to network-layer anonymization through natively integrated onion routing in consensus protocols~\cite{poscopor}.
Mandatory, optional, and TEE-based privacy have different trust and deployment profiles.

\subsubsection{Exchange Mediated and Cross Chain Obfuscation.}
Movement through centralized exchanges or conversion venues can interrupt tracing by translating funds into a different asset graph or an internal off-chain ledger~\cite{crossChainSecurityAnalysis,surveyofcrosschainprotocols,AugustoSecurityAndPrivacy}.
Large user bases supply cover traffic, but the protection applies only to external observers because the operator retains the most complete view.
This is therefore a limited form of mixing. To avoid such reliance, decentralized zero-knowledge bridges like PPBridge~\cite{ppbridge2025} enable privacy-preserving cross-chain transfers.

\subsection{Cross-Cutting Privacy Primitives}
\label{sec:primitives}

The families use a small set of recurring primitives.
Aggregation, batching, and peeling dominate custodial services and exchanges, leaving structural graph patterns~\cite{moneyLaunderingInquiry,SecureSystems,TIPPE2025301876}.
Joint transactions and shuffling create local ambiguity, but weaken with small sets, unequal amounts, or disrupted rounds~\cite{CJoin,coinShuffle,cParty}.
Commitment notes and zero-knowledge proofs provide cryptographic unlinkability in account-based systems but remain subject to public denomination, relayer, and timing metadata~\cite{TornadoCash,zcash}.
Ring signatures, stealth addresses, and confidential transactions make routine transactions contribute cover traffic, but decoy quality, wallet rules, and network observations still matter~\cite{Monero}. Recent designs also propose zkSNARK-verified hash-commitments as an efficient alternative to traditional ECDH-based stealth address derivation~\cite{zkstealth2025}.
Trusted execution hides state and ordering in hardware but adds attestation, rollback, and side channel assumptions~\cite{IntelSGX,SGXattack,brasser2017dr,biondo2018guard,matetic2017rote,cloosters2020teerex,Borrello2022AEPIC}.
Delays, fresh addresses, relayers, and network defenses such as Dandelion++~\cite{dandelionpp,randomFees} or native onion routing at the consensus layer~\cite{poscopor} do not provide privacy alone.
They protect the main primitive from simple heuristics, and they are the layer at which theoretical and practical privacy most often diverge.

\section{Comparative Review of Representative Systems}
\label{sec:systems}

This section reviews representative systems from each architectural family, focusing on the design choices that affect practical privacy, the assumptions that support them, and recurring deployment failures.
\autoref{tab:comparison} summarizes the main properties; additional detail is provided in the appendix.

\subsection{Custodial Mixers}

Custodial mixers accept funds into operator-controlled addresses and later redistribute them from pooled reserves.
They require little coordination, but the operator or a party that compromises it can reconstruct the mapping.
Pooling and redistribution also leave graph motifs such as batching and peeling chains~\cite{moneyLaunderingInquiry,SecureSystems,TIPPE2025301876}.

\paragraph{Helix and Bitcoin Fog.} Both pooled user funds, redistributed value across internal addresses, and used delays, batching, and several transfer hops.
Investigators can still exploit peeling structures, aggregate timing, and service habits~\cite{TIPPE2025301876}.
Privacy also depends on operator secrecy.
Ground truth data show how much visibility remains in the service layer~\cite{MixerServicesSecurityAnalysis}.
Helix's operator pleaded guilty, while the Bitcoin Fog operator was arrested and charged~\cite{HelixBust,bfog}.

\paragraph{Obscuro.} Obscuro combines mixing with trusted execution.
It hides ordering and mapping decisions within a TEE instead of visible on-chain coordination~\cite{obscuroPaper}.
This reduces public exposure, but privacy depends on enclave correctness, the absence of sensitive logs, and resistance to TEE compromise.
Forensic analysis of a related test environment illustrates the sensitivity of logs and network artifacts~\cite{ObscuroWasabiForensics}.
Obscuro changes rather than removes the trust requirement.

\subsection{Coordinated Bitcoin Mixers}

These protocols keep funds under user control until a joint transaction is assembled.
Privacy arises from equal outputs or cryptographic shuffling, but interactive rounds introduce liveness and Sybil surfaces.

\paragraph{CoinJoin and Wasabi Wallet.} The original CoinJoin proposal combines inputs and commonly uses equal outputs so that a passive observer cannot identify the matching output~\cite{CJoin}.
Later constructions support arbitrary values through additional cryptographic machinery~\cite{BetterCJoin}.
Wasabi Wallet implements coordinated CoinJoin and has changed its protocol over time.
Its current rounds support variable amounts and blinded credentials~\cite{wasabiwal}.
Ecosystem-level analysis shows that a small number of successor coordinators, led by kruw.io, now carry most Wasabi 2.x liquidity, with tens of concurrently active participants providing the immediate cover set~\cite{svenda2026coinjoinecosystem}.
Post-mix consolidation further narrows the nominal anonymity set by 10 to 50\% within the first day, although the loss stabilizes within a year~\cite{gavenda2025coinjoinmappings}.
Effective anonymity still depends on round composition, coordination, admission control, and disruption resistance.

\paragraph{CoinShuffle and CoinParty.} CoinShuffle uses layered encryption and output permutation so that participants randomize the mapping without exposing it to a central party~\cite{coinShuffle}.
CoinParty adds threshold cryptography and explicit commitments~\cite{cParty}.
TumbleBit is a contextual payment hub comparator with a formal unlinkability treatment against an actively malicious hub~\cite{tumblebit}.
These constructions remain interactive.
Unavailable users or strategic aborts can make the privacy gain costly or unattainable.

\subsection{Contract Mixers on Account Based Chains}

These systems place mixing logic in public contracts.
Users deposit and later prove withdrawal rights in zero knowledge.
No single party holds both funds and the plaintext mapping, but host ledger metadata remains visible.

\paragraph{M\"obius, AMR, and MixEth.} M\"obius uses ring signatures~\cite{morbius}.
AMR and MixEth employ note-based or zero-knowledge approaches whose privacy argument is stronger because withdrawal logic is cryptographically separated from the deposit~\cite{AMR,mixeth}.
Deposit timing, gas funding, and the account interaction graph remain observable, so unlinkability at the contract level does not automatically imply unlinkability at the user level.
Participation incentives, fee spikes, or congestion can reduce practical value even when the proof system is sound.

\paragraph{Tornado Cash.} Tornado Cash popularized fixed denomination zkSNARK pools with optional relayers~\cite{TornadoCash}.
No operator learns the mapping and the contract is publicly auditable.
User behavior, relayer patterns, and gas metadata can reduce effective sets far below advertised sizes~\cite{beres2021,TCAnalysis,Wang2023ZKPmixers}.
The United States Treasury designated Tornado Cash in August 2022, and empirical work reports a subsequent decline in deposits~\cite{ofacTornadoCash,Wang2023ZKPmixers}.
Treasury removed the designation in March 2025~\cite{ofacTornadoCashDelisting}.

\subsection{Privacy Layers}

These designs embed shielded execution or private state in an application layer above a host chain.

\paragraph{RAILGUN.} RAILGUN maintains a unified shielded UTXO pool for arbitrary tokens and variable amounts~\cite{railgun}.
It avoids fixed-denomination partitioning.
Huseynov et al. link roughly 17.65\% of withdrawals to deposits through timing, address reuse, transaction graph proximity, and knapsack matching of amounts, for a median anonymity loss of 3.42 bits~\cite{huseynov2026railgun}.
Private Proofs of Innocence provide an early privacy-preserving compliance mechanism~\cite{railgunPPOI}.

\paragraph{Aztec.} Aztec is a privacy focused Layer 2 rollup that separates public and private computation.
Private state transitions execute on clients and produce validity proofs for settlement on Ethereum~\cite{aztec}.
PLONK is a foundation for this class of proof systems~\cite{plonk}.
Aztec generalizes unlinkability to programmable private computation, but privacy still depends on adoption and host chain constraints.
The contextual Penumbra and Namada designs share the architectural dependence on aggregate shielded activity~\cite{penumbra,namada}.

\subsection{Privacy Native Systems}

These internalize unlinkability in the base protocol so ordinary activity contributes cover traffic.

\paragraph{Monero.} Ring signatures, stealth addresses, and RingCT provide privacy by default~\cite{Monero,RingCT}.
Routine outputs enlarge the pool from which later wallets select decoys.
Decoy selection, temporal heuristics, and wallet behavior can still reduce effective anonymity~\cite{MoneroAnalysis}.

\paragraph{Zcash and Mimblewimble Chains.} Zcash offers optional shielded transactions with zkSNARKs~\cite{zcash}.
Low adoption reduces cover traffic and enables heuristics at the boundary of the shielded pool~\cite{zcashAnalysis}.
Grin and Beam use Mimblewimble commitments and transaction aggregation~\cite{mimblewimble}.
Their distinct mechanism and limited comparable evidence are why they are described in the corpus but not scored as Zcash equivalents in the attack matrix.

\paragraph{Confidential execution platforms.} Oasis, Secret, Integritee, and Phala hide state and execution inside TEEs~\cite{oasis,secret,integritee,phala}.
Other architectures, such as AQUA\-REUM~\cite{aquareum}, explicitly leverage TEEs alongside public blockchains to provide centralized ledgers with privacy along verifiability, non-equivocation, and censorship-evidence.
They support private contracts but concentrate risk in hardware compromise and side channel attacks~\cite{SGXattack,Borrello2022AEPIC}.

\subsection{Exchange Mediated and Cross Chain Obfuscation}

Centralized exchanges and cross-chain venues pool deposits and convert assets, creating ambiguity for external observers.
Tracing across ledgers shows that exchange flows, conversion rules, and timing still permit analysis~\cite{crossChainSecurityAnalysis}.
The exchange retains the complete mapping.
This supplies privacy against outsiders but not against the operator.

\begin{table}[t]
    \centering
    \setlength{\tabcolsep}{3pt}
    \renewcommand{\arraystretch}{1.16}
    \scriptsize
    \caption{Representative families and dominant privacy tradeoffs.}
    \label{tab:comparison}
    \resizebox{\textwidth}{!}{%
    \begin{tabular}{@{}>{\raggedright\arraybackslash}m{1.8cm}>{\raggedright\arraybackslash}m{2.8cm}>{\raggedright\arraybackslash}m{3.4cm}>{\raggedright\arraybackslash}m{3.2cm}>{\raggedright\arraybackslash}m{3.8cm}@{}}
    \toprule
    \multicolumn{2}{c}{\textbf{Identification}} & \multicolumn{3}{c}{\textbf{Privacy profile}} \\
    \cmidrule(lr){1-2}\cmidrule(lr){3-5}
    \textbf{Family} & \textbf{Representatives} & \textbf{Main privacy lever} & \textbf{Trust anchor} & \textbf{Dominant failure mode} \\
    \midrule
    \textbf{Custodial mixers} & Helix~\cite{helix}, Bitcoin Fog~\cite{bfog}, Obscuro~\cite{obscuroPaper} & Service pooling, batching, delays, and optional TEE protection & Operator records; enclave integrity for Obscuro & Records, redistribution motifs, and TEE compromise \\
    \textbf{Coordinated Bitcoin mixing} & CoinJoin~\cite{CJoin}, CoinShuffle~\cite{coinShuffle}, CoinParty~\cite{cParty}, Wasabi Wallet~\cite{wasabiwal} & Joint transactions, output shuffling, and blinded coordination & Sufficient honest online participants and coordinator behavior & Sybil disruption, aborts, and small rounds \\
    \textbf{Ethereum contract mixers} & M\"obius~\cite{morbius}, AMR~\cite{AMR}, MixEth~\cite{mixeth}, Tornado Cash~\cite{TornadoCash} & Notes, ring signatures, zero knowledge proofs, and optional relayers & Correct contracts, proof soundness, and careful use & Timing, gas funding, relayers, and amount partitioning \\
    \textbf{Privacy layers} & RAILGUN~\cite{railgun}, Aztec~\cite{aztec} & Shielded UTXO pools, private L2 execution, variable amounts, and validity proofs & Proof soundness, sufficient cover traffic, and host chain operation & Amount fingerprints, low adoption, proving overhead, and host chain metadata \\
    \textbf{Privacy native currencies} & Monero~\cite{Monero}, Zcash~\cite{zcash}, Grin, Beam~\cite{mimblewimble} & Routine private transfers, shielded transactions, and confidential aggregation & Protocol parameters, wallet behavior, and private transaction adoption & Decoy heuristics, observable pool boundaries, and limited comparable evidence \\
    \textbf{Confidential execution platforms} & Oasis~\cite{oasis}, Secret~\cite{secret}, Integritee~\cite{integritee}, Phala~\cite{phala} & Confidential contracts and protected state within TEEs & Enclave integrity, attestation, and resistance to side channels & Hardware compromise, rollback, and limited cover traffic \\
    \textbf{Exchange mediated obfuscation} & CEX flows across chains~\cite{crossChainSecurityAnalysis,AugustoSecurityAndPrivacy} & Custodial batching and asset conversion across ledgers & Exchange records and the KYC perimeter & The operator retains attribution; records reveal the mapping \\
    \bottomrule
    \end{tabular}}
\end{table}

\begin{table}[t]
    \centering
    \setlength{\tabcolsep}{3pt}
    \renewcommand{\arraystretch}{1.07}
    \footnotesize
    \caption{Technical attack surfaces across representative systems.
    Columns map to the adversary archetypes in \autoref{sec:archetypes}.
    Cells follow the rubric in \autoref{sec:rubric}: \cmark{} resistant, $\circ$ partial, \xmark{} exposed, and \na{} not applicable.
    \autoref{sec:appendix-justifications} records the evidence basis.}
    \label{tab:attacks}
    \resizebox{\textwidth}{!}{%
    \begin{tabular}{@{}l l l c c c c c c@{}}
    \toprule
    \multirow{2}{*}{\textbf{System}} & \multirow{2}{*}{\textbf{Primary primitive}} & \multirow{2}{*}{\textbf{Ledger}} & \multicolumn{6}{c}{\textbf{Technical attack surface}} \\
    \cmidrule(lr){4-9}
     & & & \shortstack{\textbf{Operator}\\\textbf{mapping}} & \shortstack{\textbf{Timing}\\\textbf{analysis}} & \shortstack{\textbf{Sybil}\\\textbf{or DoS}} & \shortstack{\textbf{Amount}\\\textbf{linkage}} & \shortstack{\textbf{TEE}\\\textbf{compromise}} & \shortstack{\textbf{Network}\\\textbf{metadata}} \\
    \midrule
    \textbf{Helix~\cite{helix}, Bitcoin Fog~\cite{bfog}} & Pool and peel & UTXO & \ba & \pa & \gr & \pa & \na & \ba \\
    \textbf{Obscuro~\cite{obscuroPaper}} & TEE shuffle & UTXO & \pa & \pa & \gr & \pa & \ba & \ba \\
    \textbf{CoinJoin~\cite{CJoin}} & Joint transaction & UTXO & \gr & \pa & \ba & \ba & \na & \ba \\
    \textbf{CoinShuffle~\cite{coinShuffle}} & Encrypted shuffle & UTXO & \gr & \pa & \ba & \pa & \na & \ba \\
    \textbf{CoinParty~\cite{cParty}} & Threshold cryptography & UTXO & \gr & \pa & \pa & \pa & \na & \ba \\
    \textbf{Wasabi Wallet~\cite{wasabiwal}} & Coordinated CoinJoin & UTXO & \pa & \pa & \pa & \pa & \na & \pa \\
    \textbf{M\"obius~\cite{morbius}} & Ring signature & Account & \gr & \pa & \pa & \pa & \na & \ba \\
    \textbf{AMR~\cite{AMR}} & zkSNARK & Account & \gr & \pa & \pa & \pa & \na & \ba \\
    \textbf{MixEth~\cite{mixeth}} & zkSNARK & Account & \gr & \pa & \pa & \pa & \na & \ba \\
    \textbf{Tornado Cash~\cite{TornadoCash}} & zkSNARK & Account & \gr & \ba & \gr & \pa & \na & \pa \\
    \textbf{RAILGUN~\cite{railgun}} & zkSNARK & Account & \gr & \pa & \gr & \pa & \na & \ba \\
    \textbf{Aztec~\cite{aztec}} & Validity proofs & L2 & \gr & \pa & \gr & \gr & \na & \ba \\
    \textbf{Monero~\cite{Monero}} & RingCT & UTXO & \gr & \pa & \gr & \gr & \na & \pa \\
    \textbf{Zcash~\cite{zcash}} & zkSNARK & UTXO & \gr & \pa & \gr & \gr & \na & \ba \\
    \textbf{Oasis~\cite{oasis}, Secret~\cite{secret}} & TEE & Account & \gr & \pa & \gr & \gr & \ba & \ba \\
    \textbf{Integritee~\cite{integritee}, Phala~\cite{phala}} & TEE & Account & \gr & \pa & \gr & \gr & \ba & \ba \\
    \textbf{CEX flows~\cite{crossChainSecurityAnalysis,AugustoSecurityAndPrivacy}} & Custody and conversion & Multiple & \ba & \pa & \gr & \pa & \na & \ba \\
    \bottomrule
    \end{tabular}}
\end{table}

\section{Synthesis and Research Directions}
\label{sec:synthesis}

This section addresses RQ2 and RQ3.
The comparison shows that a system's privacy depends on where uncertainty is created and which actor can collapse it.
The same nominal anonymity set has different value when timing, relayers, operator records, or hardware compromise can narrow it.
The patterns below answer RQ2 by identifying which adversaries remain effective and which operational factors govern outcomes; the design lessons answer RQ3.

\subsection{Reading the Tables as a Whole}
\label{sec:crosstable}
\autoref{tab:comparison} and \autoref{tab:attacks} reveal patterns that individual descriptions do not show.
The matrix has 17 rows, six surfaces, and 102 cells.
Fourteen cells are \na{} because most rows do not use a TEE.
Of the 88 evaluable cells, 28 (31.8\%) are \cmark{}.
Operator compromise has 13 resistant ratings, Sybil disruption has ten, and amount linkage has five.
Timing, network metadata, and TEE compromise have no resistant rating.

The unweighted family counts are descriptive summaries, not privacy rankings.
Privacy native rows have six resistant ratings among ten applicable cells.
Privacy layers have five among ten, and confidential execution has six among 12.
Coordinated protocols have three among 20, contract mixers five among 20, custodial mixers two among 11, and exchange flows one among five.
The denominators differ because rows aggregate systems unevenly and TEE cells can be inapplicable.
Every row has at least one partial or exposed surface.
The appendix reports the family aggregates.
The archetypes in \autoref{sec:archetypes} explain these results.
$\mathcal{A}_{\mathrm{custodial}}$ is resisted on the operator surface by 13 rows without a plaintext operator mapping.
$\mathcal{A}_{\mathrm{chain}}$ retains timing information in every row: one is exposed and 16 are partial.
$\mathcal{A}_{\mathrm{coord}}$ is resisted in ten rows, while coordinated protocols are exposed or partial.
Optional network protections yield partial ratings for Wasabi Wallet, Tornado Cash, and Monero; the other rows specify no network decoupling.
$\mathcal{A}_{\mathrm{hw}}$ applies to three TEE rows, all of which expose protected state if the enclave is compromised.

Families therefore share predictable weaknesses.
Custodial mixers and exchange flows expose the operator surface.
TEE designs expose the hardware surface, and coordinated protocols are most affected by Sybil disruption.
Timing remains partial across all families.
Network protection is absent or optional.
Cover traffic and adoption affect privacy layers, native systems, and contract mixers.
Timing, gas funding, relayers, and denominations can narrow candidate sets.
Interactive protocols also depend on liveness and Sybil resistance.
Stronger cryptography can reduce operator or amount exposure while leaving adoption, timing, network, or hardware risks in place.
Trends across columns are summarized in \autoref{fig:attacks-summary}.
Operational factors are consolidated by family in \autoref{tab:operational-factors} in the appendix.

\begin{figure}[t]
\centering
\begin{tikzpicture}[x=0.47cm,y=0.58cm]
  \foreach \tick in {0,5,10,15} {
    \draw[gray!25] (0,0.45) -- (0,-5.65);
    \node[font=\scriptsize, anchor=north] at (\tick,-6.05) {\tick};
    \draw[gray!25] (\tick,0.45) -- (\tick,-5.65);
  }
  \foreach \lab/\n [count=\i from 0] in {Operator mapping/13,Timing/0,Sybil disruption/10,Amount linkage/5,TEE compromise/0,Network metadata/0} {
    \node[font=\scriptsize, anchor=east] at (-0.25,-\i) {\lab};
    \draw[fill=black!25,draw=black!55] (0,-\i-0.25) rectangle (\n,-\i+0.25);
    \node[font=\scriptsize\bfseries,anchor=west] at (\n+0.2,-\i) {\n/17};
  }
  \node[font=\scriptsize] at (7.5,-7.1) {Rows rated resistant};
\end{tikzpicture}
\caption{Resistant ratings by technical surface in \autoref{tab:attacks}.
Counts are unweighted author assessments, not an overall privacy score.
TEE compromise applies to three rows; all three are exposed.}
\label{fig:attacks-summary}
\end{figure}

\subsection{Interpreting the Comparative Ratings}

The ordinal ratings are not a scalar measure of overall privacy.
Each cell records a structural defense against one capability.
A resistant rating on one surface does not offset exposure on another, while a partial rating denotes dependence on an observable deployment condition.
The surfaces also compose: ledger and network observations can be combined, operator records can be correlated with public transactions, and round participants can use timing to select disruption targets.
The matrix separates these concerns.
The corpus table and appendix preserve the evidence behind its 17 rows.

\subsection{Nominal Anonymity Is a Weak Metric}

Across mixer families, nominal anonymity set size is an unreliable privacy proxy.
In CoinJoin systems, the set shrinks when outputs differ, rounds are small, or participants are adversarial.
In contract mixers, a large pool can conceal narrow timing windows or rare denominations.
Posterior uncertainty, captured by $k_{\mathrm{eff}}$ (\autoref{eq:keff}), is the appropriate model when a posterior can be estimated.
Otherwise, candidate counts should be reported as heuristic bounds.
A formal model for distributed shuffling similarly shows that set size can overstate privacy when shuffling is partial and participants are corrupted~\cite{cinal2026shuffling}.

\paragraph{Worked example: Tornado Cash under behavioral heuristics.}
Consider the Tornado Cash 100~ETH pool studied by Wang et al.~\cite{Wang2023ZKPmixers} and shown in \autoref{fig:funnel}.
At their observation point, it contained 30{,}301 deposits from 6{,}433 unique depositors; 5{,}608 had a positive balance.
These values are heuristic candidate counts, not estimates of $k_{\mathrm{eff}}$.
Across the Ethereum pools studied, the authors' combined behavioral heuristics reduced the observed anonymity set by 27.34\% and increased simple linking success by 37.63\%.
Tang et al.~\cite{TCAnalysis} also identify timing structure between deposits and withdrawals.
The Treasury sanctioned Tornado Cash in 2022~\cite{ofacTornadoCash}; Wang et al. report a subsequent decrease in deposits.

\begin{figure}[b]
\centering
\begin{tikzpicture}[
  bar/.style={draw=black!55, align=center, font=\scriptsize, rounded corners=1.2pt, minimum height=5mm, inner sep=1.5pt},
  stageone/.style={bar, text width=2.95cm},
  stagetwo/.style={bar, text width=2.55cm},
  stagethree/.style={bar, text width=2.15cm},
  stagefour/.style={bar, text width=1.75cm},
  sys/.style={font=\scriptsize\bfseries, align=center, text width=3.2cm},
  arr/.style={-{Stealth[length=1.1mm]}, semithick, gray!65},
]
\node[sys] (t0) at (0,0) {Tornado Cash};
\node[stageone, fill=blue!8, below=0.7mm of t0] (t1) {30{,}301 observed deposits};
\node[stagetwo, fill=blue!16, below=0.7mm of t1] (t2) {6{,}433 unique depositors};
\node[stagethree, fill=blue!25, below=0.7mm of t2] (t3) {5{,}608 active candidates};
\node[stagefour, fill=blue!36, below=0.7mm of t3] (t4) {27.34\% mean reduction};
\node[sys] (m0) at (3.75,0) {Monero w/o.\ RingCT};
\node[stageone, fill=green!8, below=0.7mm of m0] (m1) {Nominal ring with decoys};
\node[stagetwo, fill=green!16, below=0.7mm of m1] (m2) {62\% of inputs deducible};
\node[stagethree, fill=green!25, below=0.7mm of m2] (m3) {Newest input heuristic};
\node[stagefour, fill=green!36, below=0.7mm of m3] (m4) {About 80\% success};
\node[sys] (r0) at (7.5,0) {RAILGUN};
\node[stageone, fill=orange!10, below=0.7mm of r0] (r1) {Variable amount shielded pool};
\node[stagetwo, fill=orange!20, below=0.7mm of r1] (r2) {Five heuristics};
\node[stagethree, fill=orange!32, below=0.7mm of r2] (r3) {17.65\% uniquely linked};
\node[stagefour, fill=orange!45, below=0.7mm of r3] (r4) {Median loss: 3.42 bits};
\begin{scope}[on background layer]
  \draw[blue!55, semithick] (t1.north west) -- (t4.south west) (t1.north east) -- (t4.south east);
  \draw[green!55!black, semithick] (m1.north west) -- (m4.south west) (m1.north east) -- (m4.south east);
  \draw[orange!70!black, semithick] (r1.north west) -- (r4.south west) (r1.north east) -- (r4.south east);
\end{scope}
\draw[arr] (t1)--(t2); \draw[arr](t2)--(t3); \draw[arr](t3)--(t4);
\draw[arr] (m1)--(m2); \draw[arr](m2)--(m3); \draw[arr](m3)--(m4);
\draw[arr] (r1)--(r2); \draw[arr](r2)--(r3); \draw[arr](r3)--(r4);
\end{tikzpicture}
\caption{Published reductions from nominal anonymity in three families~\cite{Wang2023ZKPmixers,MoneroAnalysis,huseynov2026railgun}.
The Monero results predate mandatory RingCT.
Tornado Cash values are candidate counts and heuristic reductions, not estimates of $k_{\mathrm{eff}}$.} 
\label{fig:funnel}
\end{figure}

\paragraph{Cross-family view.}
The same pattern appears outside Tornado Cash, as \autoref{fig:funnel} shows.
Before mandatory RingCT, M\"oser et al.~\cite{MoneroAnalysis} found that chain reaction analysis could deduce the true input for 62\% of Monero inputs that used a decoy, and that a temporal heuristic selecting the newest ring member succeeded in about 80\% of cases.
Zcash studies identify heuristics at the boundary of the shielded pool when its use is limited~\cite{zcashAnalysis,deanonCrypto}.
For RAILGUN, five behavioral heuristics uniquely link 17.65\% of withdrawals, and knapsack matching produces a median anonymity loss of 3.42 bits out of a nominal 42 bits of baseline entropy~\cite{huseynov2026railgun}.
For coordinator-based CoinJoin, enumerating all input-output mappings shows that candidates are not equally likely, so raw counts overstate effective anonymity, and post-mix consolidation narrows the nominal set by 10 to 50\% within the first day~\cite{gavenda2025coinjoinmappings}.
Ecosystem measurements likewise show that concurrent participation, now concentrated under a single successor coordinator, bounds the immediate cover set more tightly than nominal pool size suggests~\cite{svenda2026coinjoinecosystem}.
Mixer evaluations should therefore distinguish pool size, heuristic candidate counts, and posterior effective anonymity, and state which side information informs each metric.

\subsection{Trust, Usability, and Cross Chain Movement}

Mixer architectures relocate trust to the operator, peers, contracts and relayers, protocol parameters, or hardware.
A contract without custody can still leak metadata, and a platform that uses a TEE can still fail under hardware compromise.
Designs should state their trust anchor and distinguish resistance to the operator from resistance to an outside observer.

Usability and liveness also affect privacy.
Interactive protocols need simultaneous participants and must tolerate churn and aborts.
Contract mixer users can weaken privacy by withdrawing quickly, reusing infrastructure, or behaving predictably.
Strong designs therefore pair cryptographic protection with accountable coordination, admission that limits Sybil participation, and guidance that avoids simple heuristics.
Movement across chains is not equivalent to privacy.
It changes the graph observed by an external analyst but does not remove the exchange operator's view~\cite{crossChainSecurityAnalysis}.

\subsection{Mixer Detection and Forensic Evolution}

The adversarial landscape changes as forensics, graph analysis, and machine learning improve detection and partial deanonymization~\cite{howToSpotAMixer,AIdetection,KumaresanTracingSurvey2025,undesrtBCOINMixServ,everythingToKnowAboutMixers}.
Distinctive footprints, such as deposits in fixed denominations or custodial peeling chains, are easier to detect than private transactions that resemble ordinary activity.

\subsection{Evaluation and Reporting Guidance}

The comparison suggests four reporting obligations for future mixer evaluations, which separate a protocol claim from a deployment claim.
First, report the candidate population: state the observation period, the asset or denomination, the active participants, and the rules that exclude inactive or repeated activity, and describe the result as a candidate count when a posterior distribution cannot be justified.
Second, state the adversarial evidence: identify the adversary's observations, auxiliary information, and permitted actions, and distinguish a demonstrated linkage method from a structural exposure inferred from the architecture.
Third, measure the operational context: participation volume, time between related events, relayer use, fee funding, wallet defaults, and network behavior often determine whether a mechanism has usable cover traffic.
Fourth, support reproducibility: describe the data source, extraction date, filtering rules, and heuristics used to construct links, and state which results depend on unavailable service records.

\subsection{Design Lessons and Research Directions}

\paragraph{\textbf{Make privacy routine.}} Privacy should be pooled rather than exceptional.
A privacy mechanism used only by a small subgroup creates a narrow and recognizable population.
Default private transfers, regular transaction patterns, and sufficient participation provide a more stable basis for anonymity than an exceptional mixing event.

\paragraph{\textbf{Treat metadata as protocol data.}} Gas funding, relayers, round admission, note lifetimes, and network propagation are observable parts of the system.
A design should specify which metadata it leaves visible and which practices it expects from users, and evaluations should report privacy after realistic heuristics rather than after an idealized execution of the cryptographic core.

\paragraph{\textbf{State residual trust boundaries.}} Every family retains an authority or assumption, whether an operator, a quorum of participants, a proof system, a relayer set, a wallet implementation, or an enclave.
A clear trust boundary distinguishes privacy against an external observer from privacy against the actor that operates the service.

\paragraph{\textbf{Use accountable compliance mechanisms.}} Compliance mechanisms should constrain narrowly defined and auditable predicates rather than grant broad operator discretion.
Selective disclosure can allow a user to demonstrate a required property without revealing a complete transaction history, provided that its governance rules, admissibility conditions, and failure modes are explicit. 
%Recent frameworks show how zero-knowledge proofs and secure multiparty computation enable collaborative anti-money laundering (AML) compliance without exposing sensitive data~\cite{femiak2025zkp}.
Legal constraints continue to change.
The Fifth Circuit limited the sanctioning of immutable Tornado Cash contracts, and Treasury removed the designation in March 2025~\cite{vanLoon2024,ofacTornadoCashDelisting}.
Policy recognizes lawful privacy uses while enforcement against mixer operators continues~\cite{treasuryGenius2026,samouraiDOJ2024}.
From 10 July 2027, Article~79 of the EU Anti-Money Laundering Regulation will restrict anonymizing accounts at regulated institutions and crypto asset service providers~\cite{euAMLR2024}.
Transaction controls can also be evaded by hidden protocol layers~\cite{cinal2025anamorphic}.
RAILGUN's Private Proofs of Innocence and zkMixer illustrate selective disclosure, but governance remains unresolved~\cite{railgunPPOI,zkMixer2025}.
Privacy layers such as Aztec~\cite{aztec} also support a future in which privacy is an execution property rather than a mixer overlay.
This requires efficient proof generation on client devices, private state management, and better integration with network anonymity~\cite{dandelionpp}.
Hybrid designs that pair shielded proofs with trusted execution extend private computation but relocate rather than remove the hardware attack surface~\cite{cinal2025auntie}.

\section{Conclusion}
\label{sec:conclusion}

A cryptographic mechanism alone cannot establish the privacy of a mixing system.
An advertised anonymity set is also insufficient.
Practical privacy emerges from the complete deployment, including participation volume, transaction timing, amount patterns, relayer use, network exposure, operator records, wallet behavior, and hardware assumptions.
Protection at one layer can remain ineffective when these surrounding conditions narrow the plausible transaction links.

The comparison leads to a broader design principle: privacy is strongest when it is routine, pooled across ordinary activity, and supported by explicit trust boundaries.
Future systems should report privacy against concrete observations and known heuristics, treat metadata as part of the protocol model, and reduce exceptional behavior that makes private transactions easy to recognize.
Compliance mechanisms should prove narrowly defined properties without granting an operator broad access to transaction histories.
These conclusions provide a common basis for comparing current architectures and for designing blockchain privacy systems whose operational guarantees more closely match their cryptographic claims.

\subsubsection*{Acknowledgments.}
This work was supported by the Brno University of Technology internal project FIT-S-26-9011 and Science Grant Agency -- project VEGA
1/0771/26, and by the Slovak Research and Development Agency under the Contract no. APVV-25-0804. 
The authors acknowledge the use of generative AI tools for assistance with grammar and text refinements.
The authors reviewed and edited the text, and take full responsibility for the content.

\subsubsection*{Disclosure of Interests.}
The authors have no competing interests relevant to the content of this article.

\newpage
\bibliographystyle{splncs04}
\bibliography{software}

@String{Computing = "Computing" }

@String{Computer = "{IEEE} Computer" }

@String{Springer = "Springer-Verlag" }

@inproceedings{undesrtBCOINMixServ,
author = {Wu, Lei and Hu, Yufeng and Zhou, Yajin and Wang, Haoyu and Luo, Xiapu and Wang, Zhi and Zhang, Fan and Ren, Kui},
title = {Towards Understanding and Demystifying {Bitcoin} Mixing Services},
year = {2021},
isbn = {9781450383127},
publisher = {Association for Computing Machinery},
address = {New York, NY, USA},
url = {https://doi.org/10.1145/3442381.3449880},
doi = {10.1145/3442381.3449880},
booktitle = {Proceedings of the Web Conference 2021},
pages = {33–44},
numpages = {12},
location = {Ljubljana, Slovenia},
series = {WWW '21}
}

@inproceedings{everythingToKnowAboutMixers,
  title={Everything you ever wanted to know about {Bitcoin} mixers (but were afraid to ask)},
  author={Pakki, Jaswant and Shoshitaishvili, Yan and Wang, Ruoyu and Bao, Tiffany and Doup{\'e}, Adam},
  booktitle={Financial Cryptography and Data Security: 25th International Conference, FC 2021, Virtual Event, March 1--5, 2021, Revised Selected Papers, Part I 25},
  pages={117--146},
  year={2021},
  organization={Springer}
}

@inproceedings {MixerServicesSecurityAnalysis,
author = {Fieke Miedema and Kelvin Lubbertsen and Verena Schrama and Rolf van Wegberg},
title = {Mixed Signals: Analyzing {Ground-Truth} Data on the Users and Economics of a {Bitcoin} Mixing Service},
booktitle = {32nd USENIX Security Symposium (USENIX Security 23)},
year = {2023},
isbn = {978-1-939133-37-3},
address = {Anaheim, CA},
pages = {751--768},
url = {https://www.usenix.org/conference/usenixsecurity23/presentation/miedema},
publisher = {USENIX Association},
month = aug
}

@misc{CJoin,
  author = {Maxwell, Gregory},
  title = {{CoinJoin}: {Bitcoin} Privacy for the Real World},
  year = {2013},
  howpublished = {Bitcointalk forum post},
  url = {https://bitcointalk.org/index.php?topic=279249.0}
}

@inproceedings{sybilAttack,
  title={The {Sybil} attack},
  author={Douceur, John R},
  booktitle={International workshop on peer-to-peer systems},
  pages={251--260},
  year={2002},
  organization={Springer}
}

@misc{HelixBust,
  title = {Ohio Resident Pleads Guilty to Operating Darknet-Based {Bitcoin} ``Mixer'' That Laundered Over \$300 Million},
  author = {{Office of Public Affairs, U.S. Department of Justice}},
  year = {2021},
  howpublished = {\url{https://www.justice.gov/archives/opa/pr/ohio-resident-pleads-guilty-operating-darknet-based-bitcoin-mixer-laundered-over-300-million}},
  note = {Accessed: 2026-04-15}
}

@inproceedings {crossChainSecurityAnalysis,
author = {Haaroon Yousaf and George Kappos and Sarah Meiklejohn},
title = {Tracing Transactions Across Cryptocurrency Ledgers},
booktitle = {28th USENIX Security Symposium (USENIX Security 19)},
year = {2019},
isbn = {978-1-939133-06-9},
address = {Santa Clara, CA},
pages = {837--850},
url = {https://www.usenix.org/conference/usenixsecurity19/presentation/yousaf},
publisher = {USENIX Association},
month = aug
}

@article{mixerHistory,
  title={Untraceable electronic mail, return addresses, and digital pseudonyms},
  author={Chaum, David L},
  journal={Communications of the ACM},
  volume={24},
  number={2},
  pages={84--90},
  year={1981},
  publisher={ACM New York, NY, USA}
}

@inproceedings{moserTaint,
  title={Towards risk scoring of {Bitcoin} transactions},
  author={M{\"o}ser, Malte and B{\"o}hme, Rainer and Breuker, Dominic},
  booktitle={Financial Cryptography and Data Security: FC 2014 Workshops, BITCOIN and WAHC 2014, Christ Church, Barbados, March 7, 2014, Revised Selected Papers 18},
  pages={16--32},
  year={2014},
  organization={Springer}
}

@inproceedings{coinShuffle,
  title={{CoinShuffle}: Practical decentralized coin mixing for {Bitcoin}},
  author={Ruffing, Tim and Moreno-Sanchez, Pedro and Kate, Aniket},
  booktitle={Computer Security-ESORICS 2014: 19th European Symposium on Research in Computer Security, Wroclaw, Poland, September 7-11, 2014. Proceedings, Part II 19},
  pages={345--364},
  year={2014},
  organization={Springer}
}

@InProceedings{SecureSystems,
author="de Balthasar, Thibault
and Hernandez-Castro, Julio",
editor="Lipmaa, Helger
and Mitrokotsa, Aikaterini
and Matulevi{\v{c}}ius, Raimundas",
title="An Analysis of {Bitcoin} Laundry Services",
booktitle="Secure IT Systems",
year="2017",
publisher="Springer International Publishing",
address="Cham",
pages="297--312",
isbn="978-3-319-70290-2"
}

@inproceedings{randomFees,
  author    = {Bonneau, Joseph and Narayanan, Arvind and Miller, Andrew and Clark, Jeremy and Kroll, Joshua A. and Felten, Edward W.},
  title     = {{Mixcoin}: Anonymity for {Bitcoin} with Accountable Mixes},
  booktitle = {Financial Cryptography and Data Security (FC 2014)},
  series    = {Lecture Notes in Computer Science},
  volume    = {8437},
  pages     = {486--504},
  publisher = {Springer},
  year      = {2014},
  doi       = {10.1007/978-3-662-45472-5_31}
}

@INPROCEEDINGS{moneyLaunderingInquiry,
  author={Möser, Malte and Böhme, Rainer and Breuker, Dominic},
  booktitle={2013 APWG eCrime Researchers Summit}, 
  title={An inquiry into money laundering tools in the {Bitcoin} ecosystem},
  year={2013},
  volume={},
  number={},
  pages={1-14},
  doi={10.1109/eCRS.2013.6805780}
}

@article{surveyofcrosschainprotocols,
title = {Survey of crosschain communications protocols},
journal = {Computer Networks},
volume = {200},
pages = {108488},
year = {2021},
issn = {1389-1286},
doi = {10.1016/j.comnet.2021.108488},
url = {https://www.sciencedirect.com/science/article/pii/S1389128621004321},
author = {Peter Robinson}
}

@ARTICLE{howToSpotAMixer,
  author={Xu, Chang and Xiong, Ruting and Shen, Xiaodong and Zhu, Liehuang and Zhang, Xiaoming},
  journal={IEEE Internet of Things Journal}, 
  title={How to Find a {Bitcoin} Mixer: A Dual Ensemble Model for {Bitcoin} Mixing Service Detection}, 
  year={2023},
  volume={10},
  number={19},
  pages={17220-17230},
  doi={10.1109/JIOT.2023.3275158}
}

@inproceedings{obscuroPaper,
author = {Tran, Muoi and Luu, Loi and Kang, Min Suk and Bentov, Iddo and Saxena, Prateek},
title = {Obscuro: A {Bitcoin} Mixer using Trusted Execution Environments},
year = {2018},
isbn = {9781450365697},
publisher = {Association for Computing Machinery},
address = {New York, NY, USA},
url = {https://doi.org/10.1145/3274694.3274750},
doi = {10.1145/3274694.3274750},
booktitle = {Proceedings of the 34th Annual Computer Security Applications Conference},
pages = {692–701},
numpages = {10},
location = {San Juan, PR, USA},
series = {ACSAC '18}
}

@misc{IntelSGX,
      author = {Victor Costan and Srinivas Devadas},
      title = {{Intel SGX} Explained},
      howpublished = {Cryptology ePrint Archive, Paper 2016/086},
      year = {2016},
      url = {https://eprint.iacr.org/2016/086}
}

@INPROCEEDINGS{ObscuroWasabiForensics,
  author={Young, Edward Henry and Chrysoulas, Christos and Pitropakis, Nikolaos and Papadopoulos, Pavlos and Buchanan, William J},
  booktitle={2021 International Conference on Data Analytics for Business and Industry (ICDABI)}, 
  title={Evaluating Tooling and Methodology when Analysing {Bitcoin} Mixing Services After Forensic Seizure}, 
  year={2021},
  volume={},
  number={},
  pages={650-654},
  doi={10.1109/ICDABI53623.2021.9655843}
}

@INPROCEEDINGS{BetterCJoin,
  author={Maurer, Felix Konstantin and Neudecker, Till and Florian, Martin},
  booktitle={2017 IEEE Trustcom/BigDataSE/ICESS}, 
  title={Anonymous {CoinJoin} Transactions with Arbitrary Values}, 
  year={2017},
  volume={},
  number={},
  pages={522-529},
  doi={10.1109/Trustcom/BigDataSE/ICESS.2017.280}
}

@inproceedings{cParty,
author = {Ziegeldorf, Jan Henrik and Grossmann, Fred and Henze, Martin and Inden, Nicolas and Wehrle, Klaus},
title = {{CoinParty}: Secure Multi-Party Mixing of {Bitcoins}},
year = {2015},
isbn = {9781450331913},
publisher = {Association for Computing Machinery},
address = {New York, NY, USA},
url = {https://doi.org/10.1145/2699026.2699100},
doi = {10.1145/2699026.2699100},
booktitle = {Proceedings of the 5th ACM Conference on Data and Application Security and Privacy},
pages = {75–86},
numpages = {12},
location = {San Antonio, Texas, USA},
series = {CODASPY '15}
}

@article{morbius,
  title={M{\"o}bius: Trustless Tumbling for Transaction Privacy},
  author={Meiklejohn, Sarah and Mercer, Rebekah},
  journal={Proceedings on Privacy Enhancing Technologies},
  volume={2018},
  number={2},
  pages={105--121},
  year={2018},
  doi={10.1515/popets-2018-0015}
}

@inproceedings{AMR,
author = {Le, Duc V. and Gervais, Arthur},
title = {{AMR}: autonomous coin mixer with privacy preserving reward distribution},
year = {2021},
isbn = {9781450390828},
publisher = {Association for Computing Machinery},
address = {New York, NY, USA},
url = {https://doi.org/10.1145/3479722.3480800},
doi = {10.1145/3479722.3480800},
booktitle = {Proceedings of the 3rd ACM Conference on Advances in Financial Technologies},
pages = {142–155},
numpages = {14},
location = {Arlington, Virginia},
series = {AFT '21}
}

@inproceedings{zk-SNARKs,
author = {Bitansky, Nir and Canetti, Ran and Chiesa, Alessandro and Tromer, Eran},
title = {From extractable collision resistance to succinct non-interactive arguments of knowledge, and back again},
year = {2012},
isbn = {9781450311151},
publisher = {Association for Computing Machinery},
address = {New York, NY, USA},
url = {https://doi.org/10.1145/2090236.2090263},
doi = {10.1145/2090236.2090263},
booktitle = {Proceedings of the 3rd Innovations in Theoretical Computer Science Conference},
pages = {326–349},
numpages = {24},
location = {Cambridge, Massachusetts},
series = {ITCS '12}
}

@misc{wasabiwal,
  author = {{zkSNACKs}},
  title = {{Wasabi Wallet} Documentation},
  year = {2026},
  howpublished = {\url{https://docs.wasabiwallet.io/}},
  note = {Accessed: 2026-07-10}
}

@INPROCEEDINGS{Zerocoin,
  author={Miers, Ian and Garman, Christina and Green, Matthew and Rubin, Aviel D.},
  booktitle={2013 IEEE Symposium on Security and Privacy}, 
  title={Zerocoin: Anonymous Distributed E-Cash from {Bitcoin}}, 
  year={2013},
  volume={},
  number={},
  pages={397-411},
  doi={10.1109/SP.2013.34}
}

@inproceedings{SGXattack,
author = {G\"{o}tzfried, Johannes and Eckert, Moritz and Schinzel, Sebastian and M\"{u}ller, Tilo},
title = {Cache Attacks on {Intel SGX}},
year = {2017},
isbn = {9781450349352},
publisher = {Association for Computing Machinery},
address = {New York, NY, USA},
url = {https://doi.org/10.1145/3065913.3065915},
doi = {10.1145/3065913.3065915},
booktitle = {Proceedings of the 10th European Workshop on Systems Security},
articleno = {2},
numpages = {6},
location = {Belgrade, Serbia},
series = {EuroSec'17}
}

@INPROCEEDINGS{AIdetection,
  author={Subramanian, Ezhilmathi and Selvakumara Samy, S.},
  booktitle={2023 12th International Conference on Advanced Computing (ICoAC)}, 
  title={Identifying illicit transactions in {Bitcoin} Tumbler services using Supervised Machine Learning Algorithms},
  year={2023},
  volume={},
  number={},
  pages={1-8},
  doi={10.1109/ICoAC59537.2023.10249782}
}

@misc{phala,
  author       = {Yin, Hang and Zhou, Shunfan and Jiang, Jun},
  title        = {{Phala Network}: A Confidential Smart Contract Network Based on {Polkadot}},
  howpublished = {Whitepaper},
  year         = {2021}
}

@misc{TornadoCash,
  author       = {Pertsev, Alexey and Semenov, Roman and Storm, Roman},
  title        = {{Tornado Cash} Privacy Solution, Version 1.4},
  year         = {2019},
  month        = dec,
  howpublished = {Technical white paper},
  url          = {https://berkeley-defi.github.io/assets/material/Tornado%20Cash%20Whitepaper.pdf}
}

@InProceedings{TCAnalysis,
author="Tang, Yujia
and Xu, Chang
and Zhang, Can
and Wu, Yan
and Zhu, Liehuang",
editor="Lu, Wei
and Zhang, Yuqing
and Wen, Weiping
and Yan, Hanbing
and Li, Chao",
title="Analysis of Address Linkability in {Tornado Cash} on {Ethereum}",
booktitle="Cyber Security",
year="2022",
publisher="Springer Nature Singapore",
address="Singapore",
pages="39--50",
isbn="978-981-16-9229-1",
doi="10.1007/978-981-16-9229-1_3"
}

@inproceedings{helix,
author = {Hong, Younggee and Kwon, Hyunsoo and Lee, Jihwan and Hur, Junbeom},
title = {A Practical De-mixing Algorithm for {Bitcoin} Mixing Services},
year = {2018},
isbn = {9781450357586},
publisher = {Association for Computing Machinery},
address = {New York, NY, USA},
url = {https://doi.org/10.1145/3205230.3205234},
doi = {10.1145/3205230.3205234},
booktitle = {Proceedings of the 2nd ACM Workshop on Blockchains, Cryptocurrencies, and Contracts},
pages = {15–20},
numpages = {6},
location = {Incheon, Republic of Korea},
series = {BCC '18}
}

@misc{bfog,
  title = {Individual Arrested and Charged with Operating Notorious Darknet Cryptocurrency ``Mixer''},
  author = {{Office of Public Affairs, U.S. Department of Justice}},
  year = {2021},
  howpublished = {\url{https://www.justice.gov/archives/opa/pr/individual-arrested-and-charged-operating-notorious-darknet-cryptocurrency-mixer}},
  note = {Accessed: 2026-04-15}
}

@article{MoneroAnalysis,
  author  = {M{\"o}ser, Malte and Soska, Kyle and Heilman, Ethan and Lee, Kevin and Heffan, Henry and Srivastava, Shashvat and Hogan, Kyle and Hennessey, Jason and Miller, Andrew and Narayanan, Arvind and Christin, Nicolas},
  title   = {An Empirical Analysis of Traceability in the {Monero} Blockchain},
  journal = {Proceedings on Privacy Enhancing Technologies},
  volume  = {2018},
  number  = {3},
  pages   = {143--163},
  year    = {2018},
  doi     = {10.1515/popets-2018-0025}
}

@misc{RingCT,
      author = {Shen Noether},
      title = {Ring Signature Confidential Transactions for {Monero}},
      howpublished = {Cryptology {ePrint} Archive, Paper 2015/1098},
      year = {2015},
      url = {https://eprint.iacr.org/2015/1098}
}

@inproceedings{mixeth,
  author    = {Seres, Istv{\'a}n Andr{\'a}s and Nagy, D{\'a}niel A. and Buckland, Chris and Burcsi, P{\'e}ter},
  title     = {{MixEth}: Efficient, Trustless Coin Mixing Service for {Ethereum}},
  booktitle = {1st International Conference on Blockchain Economics, Security and Protocols (Tokenomics 2019)},
  series    = {OpenAccess Series in Informatics (OASIcs)},
  volume    = {71},
  pages     = {13:1--13:20},
  publisher = {Schloss Dagstuhl},
  year      = {2020},
  doi       = {10.4230/OASIcs.Tokenomics.2019.13}
}

@misc{Monero,
	author = {{Monero Project}},
	title = {Monero: Private, Secure, Untraceable},
	year = {2025},
	howpublished = {\url{https://www.getmonero.org/}},
	note = {Accessed: 2026-04-20}
}

@misc{oasis,
  author = {{Oasis Protocol Project}},
  title = {The {Oasis} Blockchain Platform},
  howpublished = {Whitepaper},
  year = {2020},
  url = {https://oasisprotocol.org/whitepaper}
}

@misc{secret,
  author       = {{Secret Network}},
  title        = {{Secret Network}: A Privacy-Preserving Secret Contract and Decentralized Application Platform},
  howpublished = {Graypaper, \url{https://scrt.network/graypaper}},
  year         = {2021}
}

@misc{integritee,
  author       = {{Integritee Network}},
  title        = {{Integritee} Lightpaper},
  howpublished = {\url{https://integritee.network}},
  year         = {2021}
}

@techreport{zcash,
  author      = {Hopwood, Daira and Bowe, Sean and Hornby, Taylor and Wilcox, Nathan},
  title       = {{Zcash} Protocol Specification},
  institution = {Electric Coin Company},
  year        = {2016},
  note        = {\url{https://zips.z.cash/protocol/protocol.pdf}}
}

@inproceedings{zcashAnalysis,
  author    = {Kappos, George and Yousaf, Haaroon and Maller, Mary and Meiklejohn, Sarah},
  title     = {An Empirical Analysis of Anonymity in {Zcash}},
  booktitle = {Proceedings of the 27th USENIX Security Symposium},
  pages     = {463--477},
  year      = {2018}
}

@misc{deanonCrypto,
  author       = {Biryukov, Alex and Feher, Daniel},
  title        = {Deanonymization of Hidden Transactions in {Zcash}},
  howpublished = {Technical report, University of Luxembourg (CryptoLux)},
  year         = {2018},
  note         = {\url{https://cryptolux.org/images/d/d9/Zcash.pdf}}
}

@inproceedings{biondo2018guard,
	title={The Guard's Dilemma: Efficient Code-Reuse Attacks Against {Intel} {SGX}},
	author={Biondo, Andrea and Conti, Mauro and Davi, Lucas and Frassetto, Tommaso and Sadeghi, Ahmad-Reza},
	booktitle={27th {USENIX} Security Symposium ({USENIX} Security 18)},
	pages={1213--1227},
	year={2018}
}

@inproceedings{brasser2017dr,
  author    = {Brasser, Ferdinand and Capkun, Srdjan and Dmitrienko, Alexandra and Frassetto, Tommaso and Kostiainen, Kari and Sadeghi, Ahmad-Reza},
  title     = {{DR.SGX}: Automated and Adjustable Side-Channel Protection for {SGX} using Data Location Randomization},
  booktitle = {Proceedings of the 35th Annual Computer Security Applications Conference (ACSAC '19)},
  pages     = {788--800},
  year      = {2019},
  doi       = {10.1145/3359789.3359809}
}

@inproceedings{matetic2017rote,
	title={{ROTE}: Rollback protection for trusted execution},
	author={Matetic, Sinisa and Ahmed, Mansoor and Kostiainen, Kari and Dhar, Aritra and Sommer, David and Gervais, Arthur and Juels, Ari and Capkun, Srdjan},
	booktitle={26th USENIX Security Symposium (USENIX Security 17)},
	pages={1289--1306},
	year={2017}
}

@article{TIPPE2025301876,
title = {Unmixing the mix: Patterns and challenges in {Bitcoin} mixer investigations},
journal = {Forensic Science International: Digital Investigation},
volume = {52},
pages = {301876},
year = {2025},
note = {{DFRWS} EU 2025 - Selected Papers from the 12th Annual Digital Forensics Research Conference Europe},
issn = {2666-2817},
  doi = {10.1016/j.fsidi.2025.301876},
url = {https://www.sciencedirect.com/science/article/pii/S2666281725000150},
author = {Pascal Tippe and Christoph Deckers}
}

@article{Arbabi2025bitcoinmixing,
author = {Arbabi, Alireza and Shojaeinasab, Ardeshir and Najjaran, Homayoun},
title = {Mixing Services in {Bitcoin} and {Ethereum} Ecosystems: A Review},
journal = {IET Blockchain},
volume = {5},
number = {1},
pages = {e70021},
  doi = {10.1049/blc2.70021},
url = {https://ietresearch.onlinelibrary.wiley.com/doi/abs/10.1049/blc2.70021},
eprint = {https://ietresearch.onlinelibrary.wiley.com/doi/pdf/10.1049/blc2.70021},
year = {2025}
}

@article{Nardelli2025paymentsurvey,
  author = {Nardelli, Matteo and De Sclavis, Francesco and Iezzi, Michela},
  title = {A Hitchhiker's Guide to Privacy-Preserving Digital Payment Systems: A Survey on Anonymity, Confidentiality, and Auditability},
  journal = {Distributed Ledger Technologies: Research and Practice},
  year = {2026},
  doi = {10.1145/3801555}
}

@inproceedings{zkMixer2025,
  author    = {Constantinides, Theodoros and Cartlidge, John},
  title     = {{zkMixer}: A Configurable Zero-Knowledge Mixer with Anti-Money Laundering Consensus Protocols},
  booktitle = {2025 7th IEEE International Conference on Decentralized Applications and Infrastructures (DAPPS)},
  pages     = {111--120},
  year      = {2025},
  doi       = {10.1109/DAPPS65174.2025.00022}
}

@inproceedings{Wang2023ZKPmixers,
author = {Wang, Zhipeng and Chaliasos, Stefanos and Qin, Kaihua and Zhou, Liyi and Gao, Lifeng and Berrang, Pascal and Livshits, Benjamin and Gervais, Arthur},
title = {On How Zero-Knowledge Proof Blockchain Mixers Improve, and Worsen User Privacy},
year = {2023},
isbn = {9781450394161},
publisher = {Association for Computing Machinery},
address = {New York, NY, USA},
url = {https://doi.org/10.1145/3543507.3583217},
doi = {10.1145/3543507.3583217},
booktitle = {Proceedings of the ACM Web Conference 2023},
pages = {2022–2032},
numpages = {11},
location = {Austin, TX, USA},
series = {WWW '23}
}

@INPROCEEDINGS{AugustoSecurityAndPrivacy,
  author={Augusto, André and Belchior, Rafael and Correia, Miguel and Vasconcelos, André and Zhang, Luyao and Hardjono, Thomas},
  booktitle={2024 IEEE Symposium on Security and Privacy (SP)}, 
  title={{SoK}: Security and Privacy of Blockchain Interoperability},
  year={2024},
  volume={},
  number={},
  pages={3840-3865},
  doi={10.1109/SP54263.2024.00255}
}

@inproceedings{cloosters2020teerex,
	title={{TeeRex}: Discovery and Exploitation of Memory Corruption Vulnerabilities in {SGX} Enclaves},
	author={Cloosters, Tobias and Rodler, Michael and Davi, Lucas},
	booktitle={29th {USENIX} Security Symposium ({USENIX} Security 20)},
	pages={841--858},
	year={2020}
}

@misc{railgun,
  author = {{RAILGUN Project}},
  title = {{RAILGUN}: On-chain Zero-Knowledge Privacy Ecosystem},
  year = {2022},
  howpublished = {\url{https://railgun.org/}},
  note = {Accessed: 2026-04-10}
}

@misc{railgunPPOI,
  author = {{RAILGUN Project}},
  title = {Private Proofs of Innocence},
  year = {2023},
  howpublished = {\url{https://docs.railgun.org/wiki/assurance/private-proofs-of-innocence}},
  note = {Accessed: 2026-04-10}
}

@misc{aztec,
  author = {{Aztec Labs}},
  title = {Aztec: A Privacy-First Zero-Knowledge Rollup on {Ethereum}},
  year = {2024},
  howpublished = {\url{https://aztec.network/}},
  note = {Accessed: 2026-04-10}
}

@misc{plonk,
  author = {Ariel Gabizon and Zachary J. Williamson and Oana Ciobotaru},
  title = {{PLONK}: Permutations over {Lagrange}-bases for Oecumenical Noninteractive arguments of Knowledge},
  howpublished = {Cryptology {ePrint} Archive, Paper 2019/953},
  year = {2019},
  url = {https://eprint.iacr.org/2019/953}
}

@misc{KumaresanTracingSurvey2025,
  author  = {Kumar, Ayush and Thing, Vrizlynn L. L.},
  title   = {A Survey of Transaction Tracing Techniques for Blockchain Systems},
  howpublished = {arXiv preprint arXiv:2510.09624},
  year    = {2025},
  url = {https://arxiv.org/abs/2510.09624}
}

@inproceedings{Borrello2022AEPIC,
	title = {{ÆPIC Leak}: Architecturally Leaking Uninitialized Data from the Microarchitecture},
	author = {Borrello, Pietro and Kogler, Andreas and Schwarzl, Martin and Lipp, Moritz and Gruss, Daniel and Schwarz, Michael},
	booktitle = {31st USENIX Security Symposium (USENIX Security 22)},
	year = {2022},
}

@inproceedings{Tramer2020RemoteSideChannel,
author = {Florian Tram{\`e}r and Dan Boneh and Kenny Paterson},
title = {Remote {Side-Channel} Attacks on Anonymous Transactions},
booktitle = {29th USENIX Security Symposium (USENIX Security 20)},
year = {2020},
isbn = {978-1-939133-17-5},
pages = {2739--2756},
url = {https://www.usenix.org/conference/usenixsecurity20/presentation/tramer},
publisher = {USENIX Association},
month = aug
}

@inproceedings{tumblebit,
  author = {Heilman, Ethan and Alshenibr, Leen and Baldimtsi, Foteini and Scafuro, Alessandra and Goldberg, Sharon},
  title = {{TumbleBit}: An Untrusted {Bitcoin}-Compatible Anonymous Payment Hub},
  booktitle = {Network and Distributed System Security Symposium (NDSS)},
  year = {2017}
}

@article{dandelionpp,
  author = {Fanti, Giulia and Venkatakrishnan, Shaileshh Bojja and Bakshi, Surya and Denby, Bradley and Bhargava, Shruti and Miller, Andrew and Viswanath, Pramod},
  title = {{Dandelion++}: Lightweight Cryptocurrency Networking with Formal Anonymity Guarantees},
  journal = {Proceedings of the ACM on Measurement and Analysis of Computing Systems},
  volume = {2},
  number = {2},
  pages = {1--35},
  year = {2018},
  doi = {10.1145/3224424}
}

@misc{penumbra,
  author = {{Penumbra Labs}},
  title = {Penumbra: A Shielded, Cross-Chain Network for Private {DeFi}},
  year = {2024},
  howpublished = {\url{https://penumbra.zone/}},
  note = {Accessed: 2026-04-15}
}

@misc{namada,
  author = {{Anoma Foundation}},
  title = {{Namada}: Multi-Asset Shielded Pool for Cross-Chain Privacy},
  year = {2024},
  howpublished = {\url{https://namada.net/}},
  note = {Accessed: 2026-04-15}
}

@misc{mimblewimble,
  author = {Poelstra, Andrew},
  title = {{Mimblewimble}},
  year = {2016},
  howpublished = {\url{https://download.wpsoftware.net/bitcoin/wizardry/mimblewimble.pdf}},
  note = {Accessed: 2026-04-15}
}

@inproceedings{beres2021,
  author = {B{\'e}res, Ferenc and Seres, Istv{\'a}n A. and Bencz{\'u}r, Andr{\'a}s A. and Quintyne-Collins, Mikerah},
  title = {Blockchain is Watching You: Profiling and Deanonymizing {Ethereum} Users},
  booktitle = {IEEE International Conference on Decentralized Applications and Infrastructures (DAPPS)},
  year = {2021},
  doi = {10.1109/DAPPS52256.2021.00013}
}

@misc{ofacTornadoCash,
  author = {{U.S. Department of the Treasury, Office of Foreign Assets Control}},
  title = {{U.S.} Treasury Sanctions Notorious Virtual Currency Mixer {Tornado Cash}},
  year = {2022},
  howpublished = {\url{https://home.treasury.gov/news/press-releases/jy0916}},
  note = {Accessed: 2026-04-15}
}

@misc{ofacTornadoCashDelisting,
  author = {{U.S. Department of the Treasury}},
  title = {{Tornado Cash} Delisting},
  year = {2025},
  month = mar,
  howpublished = {Press release SB0057, \url{https://home.treasury.gov/news/press-releases/sb0057}},
  note = {21 March 2025; accessed 2026-04-18}
}

@misc{vanLoon2024,
  author = {{U.S. Court of Appeals for the Fifth Circuit}},
  title = {{Van Loon v.\ Department of the Treasury}, 122 {F.4th} 549},
  year = {2024},
  month = nov,
  howpublished = {Decided 26 November 2024},
  note = {Held that immutable {Tornado Cash} smart-contract addresses are not ``property'' under {IEEPA}}
}

@misc{treasuryGenius2026,
  author = {{U.S. Department of the Treasury}},
  title = {Innovative Technologies to Counter Illicit Finance Involving Digital Assets},
  year = {2026},
  month = mar,
  howpublished = {Congressional report mandated by Section~9 of the {GENIUS} Act, \url{https://home.treasury.gov/system/files/246/GENIUS-Act-Illicit-Finance-Innovation-Congressional-Report-March-2026.pdf}},
  note = {6 March 2026; accessed 2026-04-18}
}

@misc{samouraiDOJ2024,
  author = {{U.S. Attorney's Office, Southern District of New York}},
  title = {Founders and {CEO} of Cryptocurrency Mixing Service Arrested and Charged with Money Laundering and Unlicensed Money Transmitting Offenses},
  year = {2024},
  month = apr,
  howpublished = {\url{https://www.justice.gov/usao-sdny/pr/founders-and-ceo-cryptocurrency-mixing-service-arrested-and-charged-money-laundering}},
  note = {24 April 2024; accessed 2026-04-18}
}

@misc{euAMLR2024,
  author = {{European Parliament and Council}},
  title = {Regulation ({EU}) 2024/1624 of 31 {May} 2024 on the prevention of the use of the financial system for money laundering or terrorist financing ({AMLR})},
  year = {2024},
  howpublished = {Official Journal of the European Union, \url{https://eur-lex.europa.eu/legal-content/EN/TXT/?uri=OJ:L_202401624}},
  note = {Article~79 prohibits {CASPs} from handling anonymity-enhancing coins; applies from 10 July 2027}
}

@techreport{pfitzmann2010terminology,
  author      = {Pfitzmann, Andreas and Hansen, Marit},
  title       = {A Terminology for Talking about Privacy by Data Minimization: Anonymity, Unlinkability, Undetectability, Unobservability, Pseudonymity, and Identity Management},
  institution = {TU Dresden and ULD Kiel},
  type        = {Technical Report},
  note        = {v0.34},
  year        = {2010},
  month       = aug,
  url         = {https://dud.inf.tu-dresden.de/literatur/Anon_Terminology_v0.34.pdf}
}

@inproceedings{serjantov2003infotheoretic,
  author    = {Serjantov, Andrei and Danezis, George},
  title     = {Towards an Information Theoretic Metric for Anonymity},
  booktitle = {Privacy Enhancing Technologies (PET 2002)},
  editor    = {Dingledine, Roger and Syverson, Paul},
  series    = {Lecture Notes in Computer Science},
  volume    = {2482},
  pages     = {41--53},
  publisher = {Springer},
  address   = {Berlin, Heidelberg},
  year      = {2003},
  doi       = {10.1007/3-540-36467-6_4}
}

@inproceedings{diaz2003measuring,
  author    = {D{\'\i}az, Claudia and Seys, Stefaan and Claessens, Joris and Preneel, Bart},
  title     = {Towards Measuring Anonymity},
  booktitle = {Privacy Enhancing Technologies (PET 2002)},
  editor    = {Dingledine, Roger and Syverson, Paul},
  series    = {Lecture Notes in Computer Science},
  volume    = {2482},
  pages     = {54--68},
  publisher = {Springer},
  address   = {Berlin, Heidelberg},
  year      = {2003},
  doi       = {10.1007/3-540-36467-6_5}
}

@book{katz2020introduction,
  author    = {Katz, Jonathan and Lindell, Yehuda},
  title     = {Introduction to Modern Cryptography},
  edition   = {3rd},
  publisher = {CRC Press},
  address   = {Boca Raton, FL},
  year      = {2020},
  isbn      = {9780815354369},
  doi       = {10.1201/9781351133036}
}

@inproceedings{clauss2006structuring,
  author    = {Clau{\ss}, Sebastian and Schiffner, Stefan},
  title     = {Structuring Anonymity Metrics},
  booktitle = {Proceedings of the Second ACM Workshop on Digital Identity Management (DIM '06)},
  pages     = {55--62},
  publisher = {ACM},
  address    = {New York, NY, USA},
  year      = {2006},
  doi       = {10.1145/1179529.1179539}
}

@misc{huseynov2026railgun,
  author       = {Huseynov, Kanan and Shahzaib, Ali and Seres, Istv{\'a}n Andr{\'a}s and Tapolcai, J{\'a}nos},
  title        = {A Tattered Cloak of Invisibility: Measuring Anonymity Loss in {Railgun} on {Ethereum}},
  howpublished = {arXiv preprint arXiv:2606.25926},
  year         = {2026},
  note         = {\url{https://arxiv.org/abs/2606.25926}}
}

@inproceedings{cinal2025anamorphic,
  author    = {Cinal, Adrian and Kubiak, Przemys{\l}aw and Kuty{\l}owski, Miros{\l}aw and Wechta, Gabriel},
  title     = {Anamorphic {Monero} Transactions: The Threat of Bypassing Anti-money Laundering Laws},
  booktitle = {Computer Security -- ESORICS 2025},
  series    = {Lecture Notes in Computer Science},
  volume    = {16054},
  pages     = {103--123},
  publisher = {Springer},
  year      = {2025},
  doi       = {10.1007/978-3-032-07891-9_6}
}

@misc{cinal2025auntie,
  author       = {Cinal, Adrian},
  title        = {Auntie: Unobservable Contracts from {Zerocash} and Trusted Execution Environments},
  howpublished = {Cryptology ePrint Archive, Paper 2025/1965},
  year         = {2025},
  note         = {\url{https://eprint.iacr.org/2025/1965}}
}

@misc{cinal2026shuffling,
  author       = {Cinal, Adrian and Sobolewski, Oliwer and Wechta, Gabriel and Zag{\'o}rski, Filip},
  title        = {Beyond Anonymity Sets: A Security Model for Distributed Shuffling in Adversarial Environments},
  howpublished = {Cryptology ePrint Archive, Paper 2026/1258},
  year         = {2026},
  note         = {\url{https://eprint.iacr.org/2026/1258}}
}

@inproceedings{gavenda2025coinjoinmappings,
  author    = {Gavenda, Jiri and Svenda, Petr and Bobon, Stanislav and Sedlacek, Vladimir},
  title     = {Analysis of Input-Output Mappings in {CoinJoin} Transactions with Arbitrary Values},
  booktitle = {Computer Security -- ESORICS 2025},
  series    = {Lecture Notes in Computer Science},
  volume    = {16056},
  pages     = {126--146},
  publisher = {Springer},
  year      = {2025},
  doi       = {10.1007/978-3-032-07901-5_7}
}

@article{svenda2026coinjoinecosystem,
  author  = {Svenda, Petr and Gavenda, Jiri and Mavroudis, Vasilios and Hicks, Chris},
  title   = {{CoinJoin} Ecosystem Insights for {Wasabi} 1.x, {Wasabi} 2.x and {Whirlpool} Coordinator-Based Privacy Mixers},
  journal = {Proceedings on Privacy Enhancing Technologies},
  volume  = {2026},
  number  = {2},
  pages   = {557--592},
  year    = {2026},
  doi     = {10.56553/popets-2026-0061}
}

@inproceedings{kostal2026hopinzero,
  title={Hopinzero: zkSNARK-Verified Privacy for On-Chain Options with Pooled Liquidity},
  author={Ko\v{s}\v{t}\'{\'a}l, Kristi\'{\'a}n and Bitarovsk{\'y}, Roman and Masti{\v{l}}ak, Luk\'{\'a}{\v{s}}},
  booktitle={2026 IEEE International Conference on Blockchain and Cryptocurrency (ICBC)},
  year={2026},
  organization={IEEE}
}

@INPROCEEDINGS{ppbridge2025,
	author={Košťál, Kristián and Rybak, Andrii},
	booktitle={2026 IEEE International Conference on Decentralized Applications and Infrastructures (DAPPS)}, 
	title={PPBridge: A Zero-Knowledge Privacy-Preserving Cross-Chain Bridge for EVM Networks}, 
	year={2026},
	volume={},
	number={},
	pages={1-10}}

@INPROCEEDINGS{zkstealth2025,
	author={Košťál, Kristián and Častven, Lukáš and Géci, Michal},
	booktitle={2026 IEEE International Conference on Decentralized Applications and Infrastructures (DAPPS)}, 
	title={ZK-SNARK Stealth Addresses: Hash-Commitment Ownership Proofs as an Alternative to ECDH-Based Address Derivation}, 
	year={2026},
	volume={},
	number={},
	pages={1-6}}

@article{aquareum,
  title={AQUAREUM: Non-Equivocating Censorship-Evident Centralized Ledger with EVM-Based Verifiable Execution using Trusted Computing and Blockchain},
  author={Homoliak, Ivan and others},
  journal={arXiv preprint arXiv:2005.13339},
  year={2020}
}

@inproceedings{poscopor,
  title={PoS-CoPOR: Proof-of-Stake Consensus Protocol with Native Onion Routing Providing Scalability and DoS-Resistance},
  author={Homoliak, Ivan and Pere{\v{s}}{\'\i}ni, Martin and Tama{\v{s}}kovi{\v{c}}, Marek and Ponek, Timotej and Hellebrandt, Luk{\'a}{\v{s}} and Malinka, Kamil},
  booktitle={2025 7th Conference on Blockchain Research \& Applications for Innovative Networks and Services (BRAINS)},
  year={2025}
}

@article{sra,
  title={The security reference architecture for blockchains: Toward a standardized model for studying vulnerabilities, threats, and defenses},
  author={Homoliak, Ivan and Venema, Simon and Hummen, Ren{\'e} and Schukat, Michael and others},
  journal={IEEE Communications Surveys \& Tutorials},
  volume={23},
  number={1},
  pages={341--390},
  year={2020},
  publisher={IEEE}
}

\appendix
\section{Appendix}
\label{sec:appendix-justifications}

\subsection{Included Systems}

\autoref{tab:corpus} identifies the 22 systems in the purposive corpus.
The technical matrix evaluates 20 of them in 17 rows.
It aggregates systems only when their mechanism and evidence basis are alike.
Grin and Beam remain in the corpus but are not assigned the Zcash ratings because their mechanism and evidence differ.

\begin{table}[b]
\centering
\scriptsize
\caption{Included systems and their treatment in the technical matrix.}
\label{tab:corpus}
\setlength{\tabcolsep}{3pt}
\renewcommand{\arraystretch}{1.05}
\begin{tabular}{@{}>{\raggedright\arraybackslash}m{2.6cm}>{\raggedright\arraybackslash}m{5.0cm}>{\raggedright\arraybackslash}m{4.1cm}@{}}
\toprule
\textbf{Family} & \textbf{Included systems} & \textbf{Matrix treatment} \\
\midrule
\textbf{Custodial mixers} & Helix, Bitcoin Fog, Obscuro & Helix and Bitcoin Fog aggregated; Obscuro separate \\
\textbf{Coordinated mixers} & CoinJoin, CoinShuffle, CoinParty, Wasabi Wallet & One row per system \\
\textbf{Contract mixers} & M\"obius, AMR, MixEth, Tornado Cash & One row per system \\
\textbf{Privacy layers} & RAILGUN, Aztec & One row per system \\
\textbf{Privacy native} & Monero, Zcash, Grin, Beam & Monero and Zcash scored; Grin and Beam described only \\
\textbf{Confidential execution} & Oasis, Secret, Integritee, Phala & Oasis with Secret; Integritee with Phala \\
\textbf{Exchange mediated} & Centralized exchange flows across chains & One CEX flow row \\
\bottomrule
\end{tabular}
\end{table}

\subsection{Positioning Against Related Surveys}

\autoref{tab:related-surveys} situates this work against recent surveys and empirical analyses.
The comparison highlights the contribution of a common adversary model, explicit deployment evidence, and coverage across architectural families.

\begin{table}[ht]
\centering
\setlength{\tabcolsep}{2.5pt}
\renewcommand{\arraystretch}{1.05}
\scriptsize
\caption{Positioning against related surveys and analyses.}
\label{tab:related-surveys}
\begin{tabular}{@{}>{\raggedright\arraybackslash}m{2.2cm}>{\raggedright\arraybackslash}m{1.45cm}>{\raggedright\arraybackslash}m{4.0cm}>{\raggedright\arraybackslash}m{4.0cm}@{}}
\toprule
\textbf{Work} & \textbf{Focus} & \textbf{Strength} & \textbf{Gap addressed here} \\
\midrule
\textbf{Arbabi et al.~\cite{Arbabi2025bitcoinmixing}} & BTC and ETH mixers & Broad catalog of mixing services & Common adversary model and broader family coverage \\
\textbf{Nardelli et al.~\cite{Nardelli2025paymentsurvey}} & Private payments & Anonymity, confidentiality, and auditability & Mixer specific deployment failures \\
\textbf{Wang et al.~\cite{Wang2023ZKPmixers}} & ZKP mixers & Empirical anonymity analysis & Comparison across architectural families \\
\textbf{Tang et al.~\cite{TCAnalysis}} & Tornado Cash & Concrete clustering heuristics & Cross system evaluation \\
\textbf{Miedema et al.~\cite{MixerServicesSecurityAnalysis}} & Custodial mixer & Ground truth economic data & Coverage beyond one service \\
\textbf{Tippe and Deckers~\cite{TIPPE2025301876}} & Mixer forensics & Traceability patterns & Design taxonomy and adversary model \\
\midrule
\textbf{This work} & Cross family & Adversary centered taxonomy and 22-system corpus & Operational gap between mechanism and deployment \\
\bottomrule
\end{tabular}
\end{table}

\subsection{Evidence Notes for the Technical Matrix}

The notes below explain assignments that require empirical or system specific evidence.
Structural assignments follow the decision criteria in \autoref{tab:rubric}.
A network rating of \xmark{} means that the reviewed design specifies no mandatory separation between network identity and ledger activity.
A rating of $\circ$ means that the deployment offers an optional relayer, integrated proxy, or propagation defense.
This distinction follows known network inference attacks and defenses~\cite{Tramer2020RemoteSideChannel,dandelionpp}.

\noindent\textbf{Helix and Bitcoin Fog.} Operator compromise is \xmark{} because custody and service records can recover the mapping~\cite{HelixBust,bfog,MixerServicesSecurityAnalysis}.
Amount linkage is $\circ$ because peeling chains retain amount structure~\cite{TIPPE2025301876}.

\noindent\textbf{Obscuro.} Operator compromise is $\circ$ because sensitive state resides in a TEE, while the service still controls inputs and operational handling~\cite{obscuroPaper}.
TEE compromise is \xmark{} because the trusted basis admits SGX and AEPIC attack classes~\cite{SGXattack,Borrello2022AEPIC}.

\noindent\textbf{CoinJoin, CoinShuffle, and CoinParty.} Sybil disruption is \xmark{} for the first two and $\circ$ for CoinParty~\cite{sybilAttack,cParty}.
Amount linkage is \xmark{} for basic CoinJoin when unequal values remove output ambiguity and $\circ$ for the other two constructions.

\noindent\textbf{Wasabi Wallet.} Operator view, timing, Sybil disruption, and amount linkage are $\circ$.
The results depend on coordinator behavior, participant composition, credentials, and round timing~\cite{wasabiwal,ObscuroWasabiForensics,svenda2026coinjoinecosystem,gavenda2025coinjoinmappings}.
Integrated network protections justify a partial rather than exposed network rating.

\noindent\textbf{M\"obius, AMR, and MixEth.} Timing and amount linkage are $\circ$ because account interactions remain public and outcomes depend on activity and user behavior.
Network exposure is \xmark{} because the core designs specify no mandatory network decoupling~\cite{Tramer2020RemoteSideChannel}.

\noindent\textbf{Tornado Cash.} Timing is \xmark{} because empirical studies link deposits and withdrawals with behavioral heuristics~\cite{TCAnalysis,Wang2023ZKPmixers,beres2021}.
Amount linkage is $\circ$ because fixed denominations partition the pool.
Optional relayers justify a partial network rating~\cite{TornadoCash}.

\noindent\textbf{RAILGUN and Aztec.} Timing is $\circ$ because host chain ordering exposes metadata.
RAILGUN amount linkage is $\circ$: amount fingerprints and knapsack matches uniquely link 17.65\% of withdrawals, with a median loss of 3.42 bits~\cite{huseynov2026railgun}.
Neither core design specifies mandatory network identity separation.

\noindent\textbf{Monero and Zcash.} Timing is $\circ$ because decoy selection and shielded pool boundaries admit heuristics~\cite{MoneroAnalysis,zcashAnalysis,deanonCrypto}.
Monero receives $\circ$ for network metadata because propagation defenses can reduce exposure~\cite{dandelionpp}; Zcash receives \xmark{} under the stated structural criterion.

\noindent\textbf{Oasis, Secret, Integritee, and Phala.} TEE compromise is \xmark{} because side channels and memory attacks can reveal protected state~\cite{SGXattack,biondo2018guard,Borrello2022AEPIC}.
Rollback and enclave analysis provide further attack classes~\cite{brasser2017dr,matetic2017rote,cloosters2020teerex}.
Timing is $\circ$ because settlement still exposes ordering metadata.

\noindent\textbf{CEX flows.} Operator compromise is \xmark{} because an exchange retains the attribution mapping.
Amount linkage is $\circ$: conversion and batching obscure public flows, but internal ledgers preserve the relation~\cite{crossChainSecurityAnalysis,AugustoSecurityAndPrivacy}.

\subsection{Expanded Evaluation Criteria}

For operator compromise, \cmark{} requires that no operator can reconstruct the mapping, $\circ$ denotes a blinded or conditional view, and \xmark{} denotes records that reveal the mapping.
For timing, \cmark{} requires a mandatory defense against correlation, $\circ$ denotes dependence on volume or behavior, and \xmark{} denotes a successful empirical linkage method.
For Sybil disruption, \cmark{} denotes a noninteractive construction or robust admission, $\circ$ denotes dependence on participation or incentives, and \xmark{} denotes failure under strategic aborts.

For amount linkage, \cmark{} requires hidden or normalized amounts, $\circ$ denotes partial matching or candidate partitioning, and \xmark{} denotes visible amounts that enable direct linkage.
TEE compromise is \na{} when no enclave protects the mapping.
It is \xmark{} when enclave compromise reveals that state; no reviewed TEE system provides a documented intermediate case.
For network metadata, \cmark{} requires mandatory identity separation, $\circ$ denotes an optional or external defense, and \xmark{} denotes that the core design specifies no such defense.
These ordinal assessments do not replace a quantitative analysis of one deployment.

\subsection{Operational Factors by Family}

\autoref{tab:operational-factors} consolidates the deployment conditions that most strongly affect effective privacy.
It separates the source of cover traffic, the residual trust anchor, and the primary operational risk for each family.

\begin{table}[t]
\centering
\scriptsize
\setlength{\tabcolsep}{3pt}
\renewcommand{\arraystretch}{1.12}
\caption{Operational factors that determine effective privacy by family.}
\label{tab:operational-factors}
\begin{tabular}{@{}>{\raggedright\arraybackslash}m{1.85cm}>{\raggedright\arraybackslash}m{2.65cm}>{\raggedright\arraybackslash}m{2.2cm}>{\raggedright\arraybackslash}m{4.7cm}@{}}
\toprule
\textbf{Family} & \textbf{Pool structure} & \textbf{Trust anchor} & \textbf{Primary operational risk} \\
\midrule
\textbf{Custodial} & Operator pool; flexible values & Operator or enclave & Records, redistribution patterns, and TEE compromise \\
\textbf{Coordinated} & Peer rounds; variable or equal outputs & Peers and coordinator & Round composition, timing, and disruption \\
\textbf{Contract} & Public contract pools & Contract and relayers & Timing, gas funding, and relayers \\
\textbf{Privacy layer} & Shielded pool with variable values & Proof system and users & Host chain ordering and amount fingerprints \\
\textbf{Privacy native} & Routine confidential transfers & Protocol and wallet & Decoy quality and timing \\
\textbf{Confidential execution} & Enclave state; flexible values & TEE attestation & Side channels and rollback \\
\textbf{Exchange mediated} & Custodial batching and conversion & Exchange & Records and conversion timing \\
\bottomrule
\end{tabular}
\end{table}

\FloatBarrier

\subsection{Unweighted Family Aggregates}

\autoref{tab:family-stats} reports the matrix counts by family.
These unweighted fractions summarize the coded cells but do not rank overall privacy because family rows and applicable surfaces differ.

\begin{table}[ht]
\centering
\scriptsize
\caption{Descriptive resistance counts from \autoref{tab:attacks}.
Fractions exclude \na{} cells and are not overall privacy scores.}
\label{tab:family-stats}
\begin{tabular}{@{}>{\raggedright\arraybackslash}m{5.0cm}>{\raggedright\arraybackslash}m{1.0cm}>{\raggedright\arraybackslash}m{1.4cm}>{\raggedright\arraybackslash}m{1.5cm}>{\raggedright\arraybackslash}m{1.2cm}@{}}
\toprule
\textbf{Family} & \textbf{Rows} & \textbf{Resistant} & \textbf{Applicable} & \textbf{Fraction} \\
\midrule
\textbf{Custodial mixers} & 2 & 2 & 11 & 18\% \\
\textbf{Coordinated Bitcoin mixers} & 4 & 3 & 20 & 15\% \\
\textbf{Contract mixers} & 4 & 5 & 20 & 25\% \\
\textbf{Privacy layers} & 2 & 5 & 10 & 50\% \\
\textbf{Privacy native currencies} & 2 & 6 & 10 & 60\% \\
\textbf{Confidential execution} & 2 & 6 & 12 & 50\% \\
\textbf{Exchange mediated obfuscation} & 1 & 1 & 5 & 20\% \\
\bottomrule
\end{tabular}
\end{table}

\FloatBarrier
\subsection{System Coding}
\label{sec:appendix-coding}

\autoref{tab:lpr} makes the classification rule explicit for every corpus member.
Each row identifies the deployment locus, primary privacy primitive, and component or actor that can reconstruct the linkage under the stated model.

\begingroup
\scriptsize
\begin{longtable}{@{}>{\raggedright\arraybackslash}m{2.0cm}>{\raggedright\arraybackslash}m{2.25cm}>{\raggedright\arraybackslash}m{3.2cm}>{\raggedright\arraybackslash}m{4.15cm}@{}}
\caption{Explicit $(L,P,R)$ coding of the 22-system corpus.
$R$ states who or what can reconstruct the linkage under the model.}\label{tab:lpr}\\
\toprule
\textbf{System} & \textbf{Locus $L$} & \textbf{Primitive $P$} & \textbf{Reconstruction authority $R$} \\
\midrule
\endfirsthead
\multicolumn{4}{c}{\tablename\ \thetable\ (continued)}\\
\toprule
\textbf{System} & \textbf{Locus $L$} & \textbf{Primitive $P$} & \textbf{Reconstruction authority $R$} \\
\midrule
\endhead
\midrule
%\multicolumn{4}{r}{Continued on next page}\\
\endfoot
\bottomrule
\endlastfoot
\textbf{Helix} & Custodial service & Pooling and peeling & Service operator or holder of records \\
\textbf{Bitcoin Fog} & Custodial service & Pooling and delayed transfers & Service operator or holder of records \\
\textbf{Obscuro} & Custodial service & TEE shuffle & Enclave compromise; service controls inputs \\
\textbf{CoinJoin} & Coordinated protocol & Joint transaction & No single authority; participants and chain heuristics narrow links \\
\textbf{CoinShuffle} & Coordinated protocol & Encrypted output shuffle & Sufficiently colluding participants \\
\textbf{CoinParty} & Coordinated protocol & Threshold cryptography & Threshold coalition or disrupted round \\
\textbf{Wasabi Wallet} & Coordinated protocol & CoinJoin with blinded credentials & Coordinator metadata and participant composition \\
\textbf{M\"obius} & Contract & Ring signature & Ledger and network evidence \\
\textbf{AMR} & Contract & Note and zkSNARK & Ledger and network evidence \\
\textbf{MixEth} & Contract & Note and zkSNARK & Ledger and network evidence \\
\textbf{Tornado Cash} & Contract & Note and zkSNARK & Ledger, relayer, and network evidence \\
\textbf{RAILGUN} & Privacy layer & Shielded UTXO and zkSNARK & Ledger and amount heuristics \\
\textbf{Aztec} & Privacy layer & Private execution and validity proofs & Host chain and sequencer metadata \\
\textbf{Monero} & Base ledger & RingCT and stealth addresses & Chain and network heuristics \\
\textbf{Zcash} & Base ledger & Shielded zkSNARK transfer & Transparent boundary and chain heuristics \\
\textbf{Grin} & Base ledger & Mimblewimble & Interactive endpoints and network evidence \\
\textbf{Beam} & Base ledger & Mimblewimble & Interactive endpoints and network evidence \\
\textbf{Oasis} & Confidential execution & TEE protected contract state & Enclave compromise \\
\textbf{Secret} & Confidential execution & TEE protected contract state & Enclave compromise \\
\textbf{Integritee} & Confidential execution & TEE sidechain state & Enclave compromise \\
\textbf{Phala} & Confidential execution & TEE worker state & Enclave compromise \\
\textbf{CEX flows} & Exchange flow & Custody, batching, and conversion & Exchange operator or holder of records \\
\end{longtable}
\endgroup

The coding rule assigns a family by deployment locus.
Primitive and reconstruction authority then expose hybrid cases.
This rule keeps Obscuro in the custodial family without claiming that its operator sees the plaintext mapping under intact TEE assumptions.
TumbleBit is excluded because its primary object is a payment hub rather than a deployed ingress and egress pool.
Penumbra and Namada are contextual shielded designs that were outside the frozen system corpus.

\subsection{Extended System Notes}

\noindent\textbf{Custodial services.} A conventional service can vary fees, delays, and transfer paths, but these controls do not remove operator knowledge.
Ground truth studies and prosecutions show that records and operational habits remain decisive~\cite{MixerServicesSecurityAnalysis,HelixBust,bfog}.
Obscuro improves this model by placing keys and logs in a TEE and by checking deposits against manipulated chain views~\cite{obscuroPaper}.
The improvement relies on the enclave and does not remove service control of availability and inputs.

\noindent\textbf{Coordinated protocols.} CoinShuffle hides the output permutation with layered encryption, while CoinParty adds threshold commitments and recovery logic~\cite{coinShuffle,cParty}.
These mechanisms remove custody but add interaction.
TumbleBit provides a useful formal comparator because a malicious hub should not link payer and payee~\cite{tumblebit}.
Round composition, aborts, and network observations remain operational concerns.

\noindent\textbf{Contract mixers.} M\"obius uses ring signatures, while AMR, MixEth, and Tornado Cash separate deposits from withdrawal authority with private notes and proofs~\cite{morbius,AMR,mixeth,TornadoCash}.
Public execution retains timing, fee funding, relayer, and amount metadata.
The contract can therefore satisfy its proof statement even when surrounding behavior makes a link likely.

\noindent\textbf{Privacy layers.} RAILGUN removes fixed denomination pools through shielded UTXOs~\cite{railgun}.
Amount combinations and nearby activity still support linkage~\cite{huseynov2026railgun}.
Aztec extends private state to programmable execution~\cite{aztec}.
Both require sufficient routine activity and inherit ordering information from the host chain.
Private Proofs of Innocence illustrate how a user can prove a narrow compliance predicate without disclosing a complete history~\cite{railgunPPOI}.

\noindent\textbf{Privacy native systems.} Monero makes private transfer structure routine, but decoy selection and temporal information have reduced anonymity in earlier deployments~\cite{Monero,MoneroAnalysis}.
Zcash protects shielded transfers cryptographically, while optional use creates observable pool boundaries~\cite{zcash,zcashAnalysis}.
Grin and Beam use a distinct Mimblewimble construction and are therefore not assigned the Zcash matrix row~\cite{mimblewimble}.

\noindent\textbf{Confidential execution and exchanges.} Oasis, Secret, Integritee, and Phala support confidential state by relying on attested execution~\cite{oasis,secret,integritee,phala}.
Side channels, rollback, and memory attacks relocate privacy risk to hardware~\cite{SGXattack,Borrello2022AEPIC}.
Exchanges provide cover against an outside observer through custody and conversion, but their internal records preserve attribution~\cite{crossChainSecurityAnalysis,AugustoSecurityAndPrivacy}.

\end{document}